\documentclass[prl,aps,amssymb,twocolumn,superscriptaddress,notitlepage]{revtex4-1}
\usepackage{amsmath}
\usepackage{amssymb}
\usepackage{amsthm}
\usepackage{amsfonts}
\usepackage{listings}
\usepackage{physics}
\usepackage{enumerate}
\usepackage{latexsym}
\usepackage{psfrag}
\usepackage{bm}
\usepackage{graphicx}
\usepackage[caption=false]{subfig}
\usepackage{blkarray}
\usepackage{array}
\usepackage{color}
\usepackage[normalem]{ulem}
\usepackage[colorlinks]{hyperref}

\def\beq{\begin{equation}}
\def\eeq{\end{equation}}
\def\bal{\begin{aligned}}
\def\eal{\end{aligned}}

\begin{document}
\title{Axionic Band Topology in Inversion-Symmetric Weyl-Charge-Density Waves}

\author{Benjamin J. Wieder}
\thanks{\url{bwieder@mit.edu}}
\affiliation{Department of Physics, Massachusetts Institute of Technology, Cambridge, MA 02139, USA}
\affiliation{Department of Physics, Northeastern University, Boston, MA 02115, USA}
\affiliation{Department of Physics, Princeton University, Princeton, New Jersey 08544, USA}

\author{Kuan-Sen Lin}
\affiliation{Department of Physics and Institute for Condensed Matter Theory, University of Illinois at Urbana-Champaign, Urbana, IL, 61801-3080, USA}

\author{Barry Bradlyn}
\thanks{\url{bbradlyn@illinois.edu}}
\affiliation{Department of Physics and Institute for Condensed Matter Theory, University of Illinois at Urbana-Champaign, Urbana, IL, 61801-3080, USA}

\date{\today}
\begin{abstract}
In recent theoretical and experimental investigations, researchers have linked the low-energy field theory of a Weyl semimetal gapped with a charge-density wave (CDW) to high-energy theories with axion electrodynamics.  However, it remains an open question whether a lattice regularization of the dynamical Weyl-CDW is in fact a single-particle axion insulator (AXI).  In this work, we use analytic and numerical methods to study both lattice-commensurate and incommensurate minimal (magnetic) Weyl-CDW phases in the mean-field state.  We observe that, as previously predicted from field theory, the two inversion- ($\mathcal{I}$-) symmetric Weyl-CDWs with $\phi = 0,\pi$ differ by a topological axion angle $\delta\theta_{\phi}=\pi$.  However, we crucially discover that \emph{neither} of the minimal Weyl-CDW phases at $\phi=0,\pi$ is individually an AXI; they are instead quantum anomalous Hall (QAH) and ``obstructed'' QAH insulators that differ by a fractional translation in the modulated cell, analogous to the two phases of the Su-Schrieffer-Heeger model of polyacetylene.  Using symmetry indicators of band topology and non-abelian Berry phase, we demonstrate that our results generalize to multi-band systems with only two Weyl fermions, establishing that minimal Weyl-CDWs unavoidably carry nontrivial Chern numbers that prevent the observation of a static magnetoelectric response.  We discuss the experimental implications of our findings, and provide models and analysis generalizing our results to nonmagnetic Weyl- and Dirac-CDWs. 
\end{abstract}

\maketitle

In condensed matter physics, one of the most important tools is low-energy field theory.  From the ${\bf k}\cdot {\bf p}$ Hamiltonian of a solid-state material, one can develop an effective action to characterize robust, and frequently topological, long-wavelength response effects~\cite{wang2013chiral,you2016response,TitusCDW,qi2008topological,essin2009magnetoelectric,wilczekaxion,zhang1989effective,halperin1993theory,son2015composite}.  However, to extrapolate from a low-energy field theory to an experimentally observable response, one must carefully complete the theory to short (UV) wavelengths -- specifically, two field theories that are identical at the ${\bf k}\cdot {\bf p}$ level may differ at large momenta, leading to distinct physical interpretations.  For example, $\mathcal{H}({\bf q}) = \sigma^{x}q_{x} + \sigma^{y}q_{y}$ can characterize one of the twofold Dirac points in a graphene-like 2D semimetal~\cite{MeleGraphene,SemenoffGraphene,GrapheneReview}, or the isolated Dirac point on the surface of a time-reversal- ($\mathcal{T}$-) symmetric 3D topological insulator (TI)~\cite{fukanemele,Fu2007,Zahid3DTI,xia2009observation}.  While $\mathcal{H}({\bf q})$ always carries the (Hall) response of a half-level $2+1$-D Chern-Simons theory~\cite{qi2008topological,essin2009magnetoelectric,fukanemele,Fu2007,LapaParity}, the total response depends on the UV completion, which either includes compensating Dirac points in 2D~\cite{qi2008topological,haldanemodel,bernevigbook,hasan2010colloquium,fradkin1,LapaParity,WittenParity,RedlichParity,JackiwParity,haldanemodel,MulliganAXI,DiracInsulator}, or the nontrivial bulk of a 3D TI~\cite{fukanemele,Fu2007,qi2008topological,LapaParity,Hughes11,DiracInsulator,wieder2018axion,HOTIFan,vanderbiltaxion,essin2009magnetoelectric,MulliganAXI,witten2016fermion}.

\begin{figure}[t]
\includegraphics[width=\columnwidth]{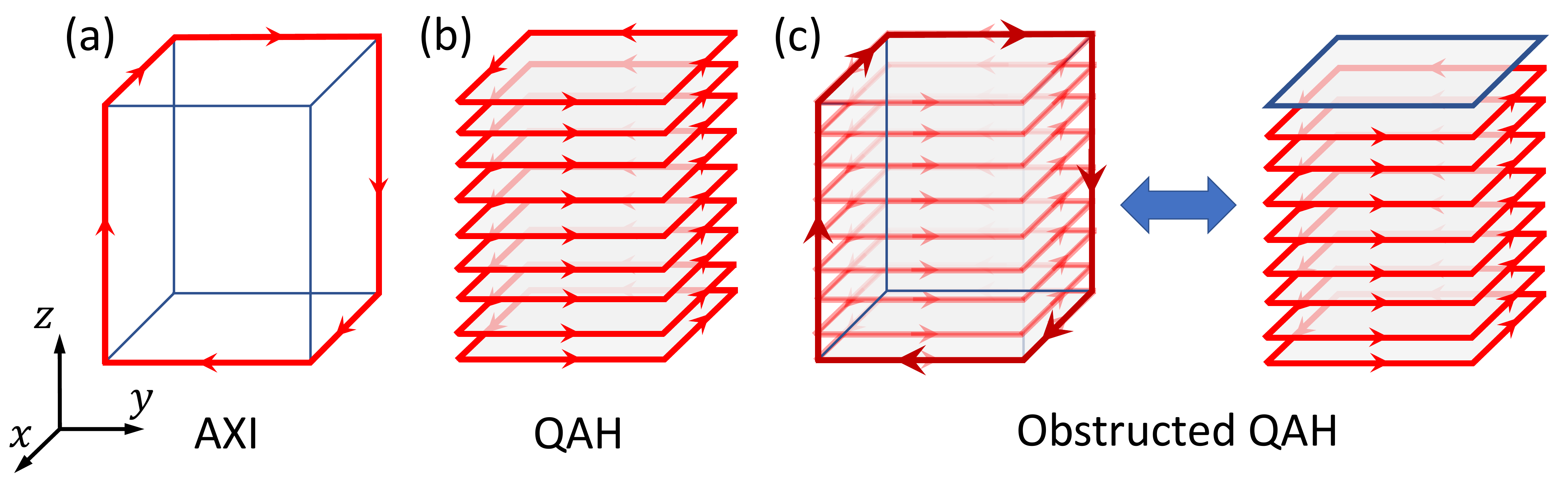}
\caption{(a) An $\mathcal{I}$-symmetric AXI with $\nu_{x,y,z}=0,\ \theta=\pi$.  (b) A $\hat{\bf z}$-directed, $\mathcal{I}$-symmetric weak stack of Chern insulators with a nontrivial QAH effect [$\nu_{z}=-1,\ \nu_{x}=\nu_{y}=\theta=0$].  (c) In a superposition of (a) and (b), the hinge states generically hybridize with the QAH surface states.  Hence, the superposition can be deformed into an oQAH insulator [$\nu_{z}=-1,\ \nu_{x}=\nu_{y}=0,\ \theta=\pi$] equivalent to shifting the QAH insulator in (b) by a half-lattice translation.  A finite-sized oQAH insulator either exhibits coexisting surface and hinge states or exactly one fewer (or one more) QAH surface state, depending on whether $\mathcal{I}$ symmetry is weakly broken.}
\label{fig:scheme}
\end{figure}

Some of the most intriguing low-energy field theories involve condensed-matter realizations of high-energy electrodynamics~\cite{wilczekaxion}.  In 3D insulators, the long-wavelength response is governed by the action:
\begin{equation}
S[A_\mu]=\frac{1}{4\pi^2}\int d^4x \epsilon^{\mu\nu\lambda\rho}(\theta\partial_\mu+ v_\mu)A_\nu\partial_\lambda A_\rho,
\label{eq:mainAction}
\end{equation}
in which ${A_\mu}$ is the electromagnetic gauge potential~\cite{qi2008topological,essin2009magnetoelectric,you2016response}~\footnote{In this work we will use Greek indices as spacetime indices, and Roman indices as purely spatial indices.}, and $v_\mu$ is a rotational-symmetry-breaking vector that determines the quantum anomalous Hall (QAH) response.  When $v_\mu=(0,\mathbf{v})$ is constant:
\begin{equation}
\sigma^H_{ij}=\frac{e^2v_k}{h}\epsilon_{kij},
\end{equation}
where $\sigma^{H}_{ij}$ is the Hall conductivity tensor.  For gapped periodic systems:
\begin{equation}
\mathbf{v} = C_{i}(\mathbf{k}\cdot \mathbf{R}_i)\mathbf{G}_i,
\end{equation}
where $\mathbf{R}_i$ is a primitive lattice vector, $\mathbf{G}_i$ is a primitive reciprical lattice vector, and $C_{i}(\mathbf{k}\cdot \mathbf{R}_i)\equiv v_i/|\mathbf{G}_i|=\nu_i$ is the \emph{weak} Chern number [Fig.~\ref{fig:scheme}(b)] in each of the Brillouin-zone- (BZ-) planes normal to $\mathbf{R}_i$~\cite{halperin1987possible,kohmoto1992diophantine,qi2008topological,haldane2004berry,you2016response,TitusCDW}~\footnote{Note that for gapless systems, such as Weyl semimetals, $C_{i}(\mathbf{k}\cdot \mathbf{R}_i)$ will instead depend on $\mathbf{k}$.}. In Eq.~(\ref{eq:mainAction}), the axion angle $\theta$ governs the magnetoelectric response~\cite{qi2008topological,essin2009magnetoelectric}; in terms of the non-abelian Berry connection ${\bf \mathcal{A}}$:
\begin{equation}
\theta[{\bf\mathcal{A}}]=\frac{1}{4\pi}\int d^3k\epsilon^{ijk}\mathrm{tr}\left(\mathcal{A}_i\partial_j\mathcal{A}_k-\frac{2i}{3}\mathcal{A}_i\mathcal{A}_j\mathcal{A}_k\right).
\label{eq:thetadef}
\end{equation}
Spatial inversion ($\mathcal{I}$) and $\mathcal{T}$ symmetries (as well as other, more complicated symmetries)~\cite{wieder2018axion} act to quantize $\theta$ as a $\mathbb{Z}_{2}$ topological invariant for which $\theta=0$ ($\theta=\pi$) is the trivial (topological) value~\cite{qi2008topological,essin2009magnetoelectric,witten2016fermion}\footnote{We note that, because Eq.~(\ref{eq:mainAction}) is an action, then only $\theta\text{ mod }2\pi$ is a gauge-invariant quantity. This is reflected in the gauge-ambiguity of the Chern-Simons term [Eq.~(\ref{eq:thetadef})].}.  In particular, $\nu_{x,y,z}=0$, $\theta=\pi$ defines a 3D TI when $\mathcal{T}$ symmetry quantizes $\theta$, and defines a magnetic axion insulator (AXI) when $\mathcal{T}$ symmetry is absent and $\theta$ is instead quantized by $\mathcal{I}$~\cite{qi2008topological,essin2009magnetoelectric,wieder2018axion,Hughes11,Turner2012,AFMTIMoore}.  In 3D TIs and AXIs, the combination of $\theta=\pi$ and $\nu_{x,y,z}=0$ leads to unusual response properties, including low-energy excitations resembling magnetic monopoles (the Witten effect~\cite{WittenEffect,rosenberg2010witten}) and quantized Faraday and Kerr rotations~\cite{qi2008topological,wu2016quantized}.  AXIs have recently been recognized as ``higher-order'' TIs~\cite{wieder2018axion,xu2020high,NaturePaper,MTQC,HOTIFan,vanderbiltaxion,benalcazar2017quantized,benalcazar2017electric,Po2017,langbehn2017reflection,watanabe2018structure,khalaf,khalaf2018higher,KoreanAXI,NicoDavidAXI2,TMDHOTI,fang2019new,JiabinZhidaAXIDirac,MurakamiAXI1,MurakamiAXI2,pozo2019quantization} (HOTIs) featuring gapped surfaces and odd numbers of sample-encircling chiral hinge modes [Fig.~\ref{fig:scheme}(a)].  AXI phases have been proposed in a number of compounds~\cite{xu2019higher,xu2020high,SarmaMobiusAXI} and observed in Mn-doped 3D TIs~\cite{AXIExp1,AXIExp2,AXIExp3,SlagerAshvinAXI}.

\begin{figure}[t]
\includegraphics[width=0.7\columnwidth]{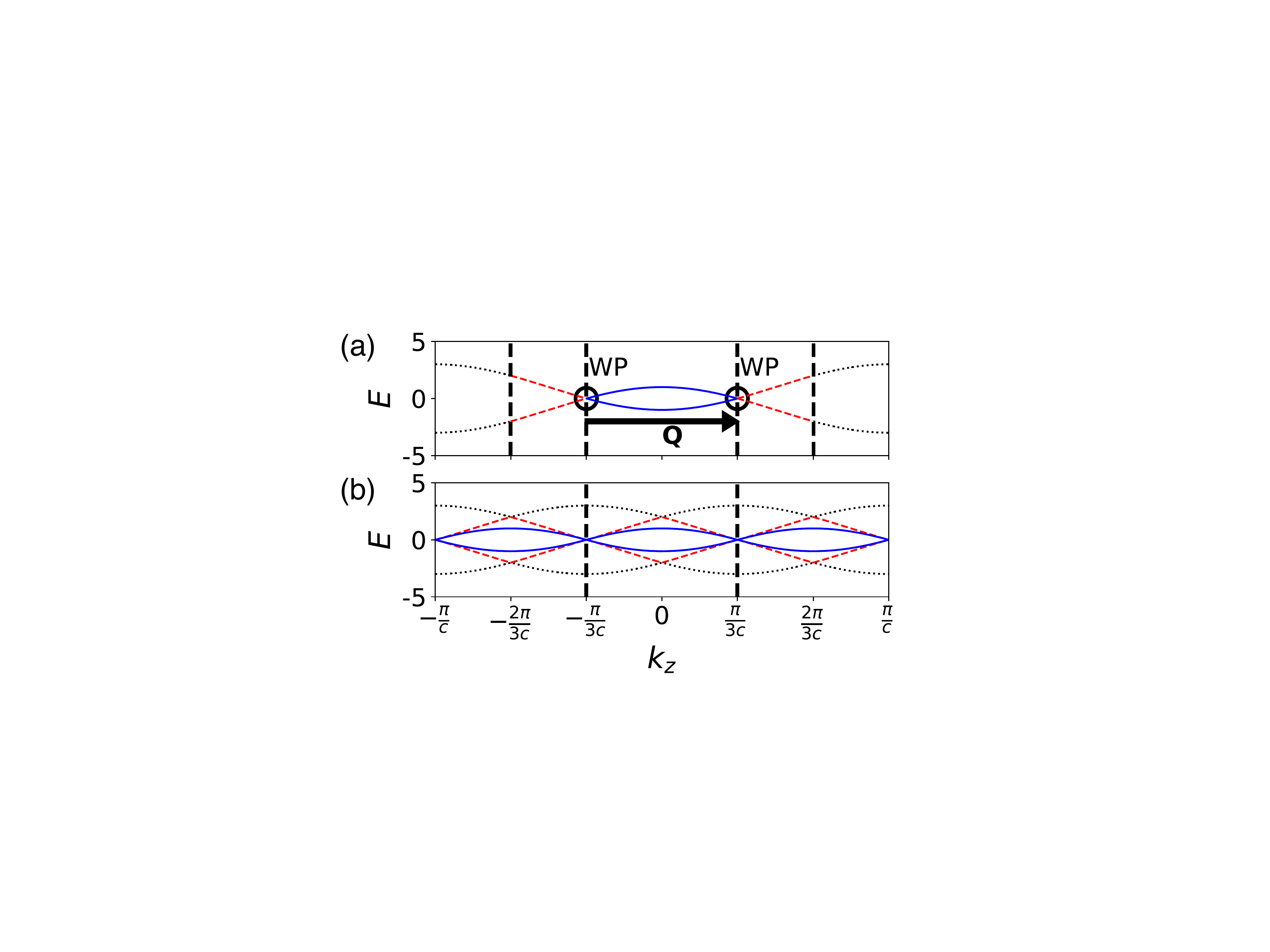}
\caption{BZ and band structure folding from a CDW with ${\bf Q}=(2\pi/Nc)\hat{\bf z}$, $N=3$. (a) A minimal $\mathcal{I}$-symmetric WSM with two WPs at $k_{z} = \pm \pi/3c$ and a BZ with ${\bf G}_{z} = (2\pi/c)\hat{\bf z}$ [Eq.~(\ref{eq:ham0})].  Solid blue (red dashed) [black dotted] bands lie in the first (second) [third] third of the first BZ.  (b) The rBZ, for which ${\bf G}_{z}' = {\bf Q}$.  The WPs from (a) are folded in (b) into an (unstable) fourfold degeneracy at the rBZ boundary.}
\label{fig:zonefold}
\end{figure}

A $\theta=\pi$ phase also appears in the low-energy field theory of a topological semimetal gapped by a charge-density wave (CDW) distortion~\cite{wang2013chiral,you2016response,BurkovCDW,zyuzin2012weyl,maciejko2014weyl}.  Specifically, in Weyl semimetals (WSMs) -- whose bulk Fermi pockets [Weyl points (WPs)] are sources and sinks of Berry curvature characterized by integer-valued topological (chiral) charges~\cite{Wan11,BalentsWeyl,armitage2018weyl} [Fig.~\ref{fig:zonefold}(a)] -- it was shown at the $\mathbf{k}\cdot\mathbf{p}$ level that $\theta({\bf x},t)=\theta_0+\phi({\bf x},t)$, where $\theta_0=\mathbf{Q}\cdot\mathbf{x}$ is the contribution to $\theta({\bf x},t)$ from two WPs separated by a momentum $\mathbf{Q}$, and $\phi({\bf x},t)$ is the (dynamical) phase of the CDW order parameter~\cite{wang2013chiral,you2016response,BurkovCDW,zyuzin2012weyl,maciejko2014weyl}. The appearance of an axionic response was attributed to the chiral anomaly in quantum field theory.  $\phi({\bf x},t)$ is a Goldstone mode, and hence can be tuned freely; uniform shifts of $\phi({\bf x},t)$ are the current-carrying sliding mode of the CDW. In disordered or incommensurate CDWs, however, $\phi({\bf x},t)$ is typically pinned to a non-universal value~\cite{fukuyama1978dynamics,bardeen1980tunneling,gruner1988dynamics}.  Refs.~\cite{wang2013chiral,you2016response,BurkovCDW,zyuzin2012weyl,maciejko2014weyl} have recently been revisited in light of experiments on the CDW compound (TaSe$_4$)$_2$I~\cite{Ta2Se8IPrepare,Cava1986} demonstrating WPs at high-temperatures~\cite{shi2019charge,OtherARPESCDWWeyl} and nonlinear negative magnetoresistance consistent with a gapped dynamical Weyl-CDW~\cite{gooth2019evidence}.

Confusingly, the dynamical Weyl-CDW is frequently labeled an AXI in the literature~\cite{wang2013chiral,you2016response}.  However, because the Weyl-CDW response in~\cite{wang2013chiral,you2016response} was derived from a ${\bf k}\cdot {\bf p}$ approximation, and at static $\phi$ in~\cite{TitusCDW}, then it remains an open and urgent question whether there exists a UV completion in which $\delta\theta_{\phi} = \theta_{\phi=\pi}-\theta_{\phi=0}\text{ mod }2\pi = \pi$ emerges due to the topology of band electrons, and whether the bulk at $\phi=0,\pi$ is a single-particle AXI.  In this work, we demonstrate that the $\mathcal{I}$-symmetric UV completion of the simplest dynamical Weyl-CDW is \emph{not} an AXI, but is instead, depending on $\phi$, one of two \emph{topologically distinct} QAH phases -- a QAH insulator or an ``obstructed'' QAH (oQAH) insulator [Fig.~\ref{fig:scheme}(b,c)] -- that differ by a fractional lattice translation.  Crucially, although $\theta$ is origin-dependent in the presence of a background QAH~\cite{qi2008topological,vanderbiltaxion,NicoDavidAXI2}, we find that the QAH and oQAH phases still \emph{differ} by an \emph{origin- (gauge-) independent, topological} axion angle $\delta\theta_{\phi}=\pi$ that reflects a difference in $\mathcal{I}$-quantized ``Chern number polarization''~\cite{NicoDavidAXI2,zilberberg2018photonic}.  This provides a direct analogy between axionic CDWs and the Su-Schrieffer-Heeger model of an $\mathcal{I}$-symmetric CDW in polyacetylene~\cite{ssh1979}, in which both phases are trivial atomic limits that differ by a fractional lattice translation corresponding to an $\mathcal{I}$-quantized topological polarization.  We demonstrate that the relative axionic response of two $\mathcal{I}$-symmetric Weyl-CDWs originates from their single-particle band topology.  We generalize our findings to multi-band systems with two WPs and to incommensurate CDWs, establishing that QAH insulators and topological phase shifts $\delta\theta_{\phi}=\pi$ are generic in minimal $\mathcal{I}$-symmetric Weyl-CDWs.  Our focus on magnetic Weyl-CDWs is further justified by recent experiments demonstrating the existence of tunable magnetic WSM phases in Co$_2$MnGa~\cite{MagneticWeylZahid}, Co$_3$Sn$_2$S$_2$~\cite{MagneticWeylYulin,MagneticWeylHaim},  CoS$_2$~\cite{CoS2LeslieMagneticWeyl}, CeAlGe~\cite{CeAlGeMagneticWeyl}, Mn$_3$Sn~\cite{Mn3SnMagneticWeyl,Mn3SnPhaseTransition}, and Mn$_3$ZnC~\cite{Mn3ZnCLeslieWeyl}, many of which host successive, symmetry-lowering magnetic phase transitions below room temperature.  In the conclusion and Supplementary Material\cite{SM}, we provide additional models and analysis generalizing our results to nonmagnetic Weyl- and Dirac-CDWs.

To begin, we introduce a simple model of a $\mathcal{T}$-broken (magnetic), $\mathcal{I}$-symmetric WSM with two WPs and with orthorhombic lattice vectors of length $a,b,c$ in the $\mathbf{\hat{x}},\mathbf{\hat{y}},\mathbf{\hat{z}}$ directions, respectively~\cite{mccormick2017minimal}:
\begin{align}
    H(\mathbf{k})&= -2[t_x\sigma^x\sin (k_xa) -t_y\sigma^y\sin (k_yb)]\nonumber \\
    & +2t_z\sigma^{z}[\cos (k_zc) -\cos\frac{Qc}{2}] \nonumber \\
    &-m\sigma^{z}[2-\cos (k_xa) - \cos (k_yb)],
\label{eq:ham0}
\end{align}
where $m/2>t_x,t_y,t_z>0$. Eq.~(\ref{eq:ham0}) is gapped at half filling at all ${\bf k}$ points away from two WPs at $\mathbf{k}=(0,0,\pm Q/2$) with chiral charges $C=\pm1$ [Fig.~\ref{fig:zonefold}(a)] related by $\mathcal{I}$, here represented by $\mathcal{I}H(\mathbf{k})\mathcal{I}^{-1} = \sigma^z H(-\mathbf{k})\sigma^z$.
As shown in~\cite{QWZ,Fang2012} and in the Supplementary Material\cite{SM}, the occupied parity ($\mathcal{I}$) eigenvalues imply the ${\bf k}$-space Chern numbers $C_{z}(ck_{z}=0)\text{ mod }2=-1$, $C_{z}(ck_{z}=\pi)\text{ mod }2=0$, mandating the appearance of the $|C|=1$ WPs.

We next construct a $\mathbf{k}\cdot\mathbf{p}$ expansion of Eq.~(\ref{eq:ham0}) about the two WPs:
\begin{equation}
H(\mathbf{q})\approx -(2t_{x}aq_x\sigma^x-2t_{y}bq_{y}\sigma^y)\tau^0 +2t_{z}cq_{z}\sin\frac{Qc}{2}\sigma^z\tau^z, 
\label{eq:kdp}
\end{equation}
where the Pauli matrices $\vec{\tau}$ act in the space of electron annihilation operators $c_{1/2,{\bf k}}$ as:
\begin{equation}
    c_\mathbf{R}\approx\sum_\mathbf{k} c_{1\mathbf{k}}e^{i{\bf R}\cdot[(Q/2)\hat{\bf z}+{\bf k}]}+c_{2\mathbf{k}}e^{-i{\bf R}\cdot[(Q/2)\hat{\bf z}-{\bf k}]}.
\label{eq:expandedop}
\end{equation}
Eq.~(\ref{eq:kdp}) can be gapped by a CDW distortion:
\begin{align}
H_{CDW}&=2\sum_{\mathbf{R}}|\Delta|\cos(QR_z+\phi)c^\dag_\mathbf{R}\sigma^z c_\mathbf{R} \\
    &=|\Delta|\sum_\mathbf{k} (c^\dag_{\mathbf{k}-\frac{Q}{2}\mathbf{\hat{z}}}\sigma^z c_{\mathbf{k}+\frac{Q}{2}\mathbf{\hat{z}}}e^{-i\phi} + \mathrm{h.c.}),
\label{eq:cdwmass}
\end{align}
which breaks the translation symmetry of Eq.~(\ref{eq:ham0}), coupling the two WPs and inducing a mass in Eq.~(\ref{eq:kdp}):
\begin{equation}
    V_\phi = |\Delta|\sigma^z(\tau^{x}\cos\phi - \tau^{y}\sin\phi),
\label{eq:kdpMass}
\end{equation}
that opens a gap at all $\phi$ for $Q\neq \pi/c$, $|\Delta|>0$.  Crucially, $\mathcal{I}$ symmetry is now represented in Eqs.~(\ref{eq:kdp}) and~(\ref{eq:kdpMass}) by $\mathcal{I}H(\mathbf{q})\mathcal{I}^{-1} = \sigma^z \tau^x H(-\mathbf{q})\sigma^z\tau^x$, such that Eq.~(\ref{eq:kdpMass}) only preserves $\mathcal{I}$ (centered at the origin) for $\phi=0,\pi$ when $Q\neq \pi/c$ (see \cite{SM}).

Consistent with previous works~\cite{wang2013chiral,you2016response,BurkovCDW,zyuzin2012weyl,maciejko2014weyl}, a domain wall between $\phi=0,\pi$ is equivalent to the critical point between a trivial insulator and an AXI~\cite{wieder2018axion,qi2008topological}.  Correspondingly, because $\{H({\bf q}),V_{\phi}\}=0$ for all $\phi$, $\mathcal{I}$-breaking defects in the space $(\Delta,\phi)$ will bind chiral modes~\cite{teo2010topological} (the axion strings in~\cite{wang2013chiral}).  In~\cite{wilczekaxion,wang2013chiral,you2016response,ryu2010topological,qi2008topological}, the authors used the chiral anomaly to motivate the appearance of chiral modes, identifying the relationship $\theta=[\pi/2](1-\mathrm{sgn}[\cos\phi])\text{ mod }2\pi$ for $\phi=0,\pi$.  $\delta \theta_{\phi} = \delta \phi$ is also consistent with magnetic symmetry-based indicators $\{\tilde{z}_4|\tilde{z}_{2x}\tilde{z}_{2y}\tilde{z}_{2z}\}$ for 3D crystals with $\mathcal{I}$ and translation symmetries~\cite{xu2020high,NaturePaper,MTQC,Po2017,watanabe2018structure,khalaf,khalaf2018higher,wieder2018axion,JiabinZhidaAXIDirac,MurakamiAXI1,MurakamiAXI2}:  
\begin{align}
\tilde{z}_4&=\frac{1}{2}\sum_{\mathbf{k}_a\in\mathrm{TRIMS}} (n^a_+- n^a_-)\text{ mod }4, \nonumber \\
\tilde{z}_{2,i} &= \frac{1}{2}\sum_{{\bf k_a\cdot R_{i}}=\pi\in\mathrm{TRIMS}}(n^a_+- n^a_-)\text{ mod }2,
\label{eq:z4def}
\end{align}
where $n^a_{\pm}$ are the number of valence $\pm 1$ parity eigenvalues at $\mathbf{k}_a$.  Specifically, $\delta \phi = \pi$ in Eq.~(\ref{eq:kdpMass}) implies that $|\delta \tilde{z}_{4}| = 2$.  However, because weak Chern numbers are only $\mathcal{I}$-symmetry-indicated modulo $2$~\cite{QWZ,Fang2012}, $|\delta\tilde{z}_4|=2$ does not itself indicate an AXI transition. Additionally, when $\nu_z\neq 0$, defining $\theta$ uniquely requires the specification of a reference state and $\mathcal{I}$ center (\emph{i.e.}, an origin)~\cite{NicoDavidAXI2}.  Furthermore, if there are other bulk or surface contributions to the topological response (\emph{e.g.}, other massive Dirac fermions at larger momenta, or a background QAH), then defects in $V_{\phi}$ will host additional states that coexist with and obscure the AXI bound states.  Therefore, to fully determine the topology of the Weyl-CDW, we will analyze the lattice-regularized UV completion [Eqs.~(\ref{eq:ham0}) and (\ref{eq:cdwmass})] beyond Eq.~(\ref{eq:z4def}).

When $Q=2\pi/Nc$, $N\in\mathbb{Z}^{+}$ in Eq.~(\ref{eq:ham0}), the CDW is lattice-commensurate, and Eq.~(\ref{eq:ham0}) remains periodic in a folded (reduced) BZ (rBZ) with ${\bf G}' = Q\hat{\bf z}$ that includes bands translated from $|k_z|> \pi/(Nc)$ [Fig.~\ref{fig:zonefold}(b)].  For all values of $N$, the two WPs fold into a linear fourfold (Dirac)~\cite{Young12} degeneracy at the rBZ boundary [Eq.~(\ref{eq:kdp})].  We deduce from the bulk parity eigenvalues that $C_z(|k_zc| < \pi/N)=-1\ ,C_z(|k_zc|>\pi/N)=0$ for all $N\in\mathbb{Z}^{+}$~\footnote{Throughout this work, we choose parameters for which $\nu_z=-1$.}, implying that $\nu_z=-1$ in the rBZ [Fig.~\ref{fig:zonefold}(b)], \emph{independent of whether $\phi=0,\pi$}.  Combining $\nu_{z}=-1$ with the ${\bf k}\cdot{\bf p}$ analysis preceding Eq.~(\ref{eq:z4def}) and fixing the origin to $z=0$ in the modulated cell, we find that $\phi=0$ [$\phi=\pi$] corresponds to a $\{2|001\}$ $\hat{\bf{z}}$-directed weak Chern (\emph{i.e.} QAH) insulator [$\{0|001\}$ oQAH insulator] (see Fig.~\ref{fig:scheme}) with $\nu_{x,y}=0$, $\nu_{z}=-1$ and $\theta=0$ [$\theta=\pi$].  Despite the QAH and oQAH insulators differing by a translation $(Nc/2)\hat{\bf z}$, $\delta\theta_{\phi}=\pi$, independent of the choice of origin.  In the SM, we provide further details, and analytically compute $\nu_{x,y,z}$ and $\delta\theta_{\phi}$ for $N=2$.

\begin{figure}[t]
\includegraphics[width=0.8\columnwidth]{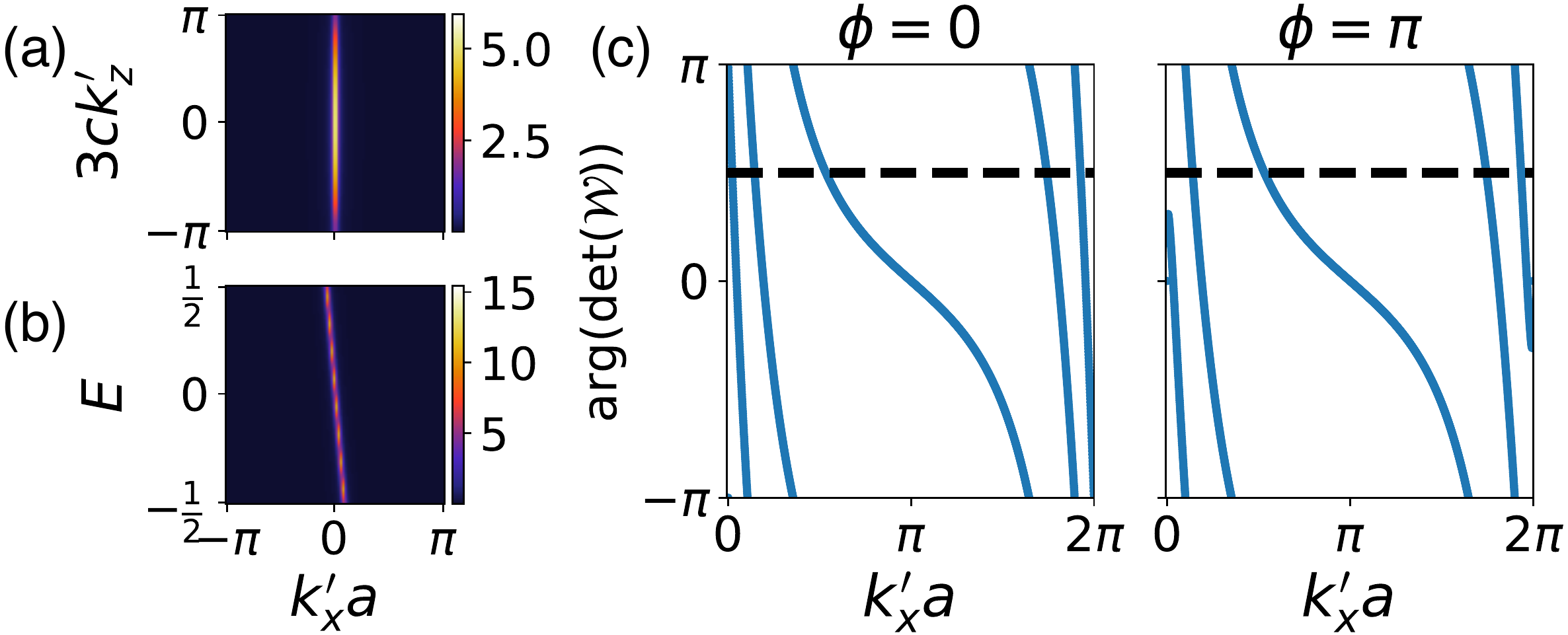}
\caption{Bulk ($\nu_{z}$) and slab ($\mathcal{G}_{z}$) Chern numbers for an $N=3$ ($Q=2\pi/3c$) commensurate CDW.  (a) The $\hat{\bf y}$-surface spectral function at $E=0$ exhibits a flat band.  (b) The $\hat{\bf y}$-surface spectral function at $k_z'=0$ exhibits $C_{z}(3k'_zc)=-1$ spectral flow along $k_{x}'$.  The spectral functions in (a,b) are the same for $\phi=0,\pi$. (c) The $\hat{\bf y}$-directed slab Berry phase~\cite{wieder2018axion} for a slab with 5 unit cells in the $\mathbf{\hat{z}}$-direction exhibits $\mathcal{G}_{z}=-5$ ($\mathcal{G}_{z}=-4$) spectral flow for $\phi=0$ ($\phi=\pi$).  (a-c) imply that the Weyl-CDW at $\phi=0$ ($\phi=\pi$) is a QAH (oQAH) insulator [Fig.~\ref{fig:scheme}(b,c)], and that $\delta\theta_{\phi}=\pi$.}
\label{fig:commensuratewilson}
\end{figure}

To explicitly determine the bulk topology, we will employ model-agnostic numerical methods~\cite{wieder2018axion,vanderbiltaxion,NicoDavidAXI2} to extract $\nu_{z}(\phi)$ and $\delta\theta_{\phi}$.  To begin, we fix the origin to the $\mathcal{I}$ center at $(x,y,z)=(0,0,0)$ in the modulated cell, and form an $\mathbf{R}_i$-directed, $\mathcal{I}$-symmetric slab. The Hall \emph{conductance} $G_{H,i}$ of the slab consists of an extensive contribution from the bulk QAH and an \emph{intensive} contribution from $\theta$ that either reflects the bulk magnetoelectric polarizability or a QAH effect offset from the origin~\cite{qi2008topological,vanderbiltaxion,NicoDavidAXI2}:
\begin{equation}
    G_{H,i}=\sigma_{H,i}L_i + (e^2\theta/h\pi),
\label{eq:thetaconductance}
\end{equation}
where $\sigma_{H,i}=e^2\nu_i\mathbf{G}_i/2\pi h$ is the Hall conductivity (given by the weak Chern number $\nu_{i}$), where $L_i$ is the (lattice-regularized) thickness of the slab.  Because a slab is a quasi-2D system, it carries a quantized Chern number $\mathcal{G}_{i}$ that is related to Eq.~(\ref{eq:thetaconductance}) by $\mathcal{G}_{i}e^2/h = G_{H,i}$.  For $\mathcal{I}$-symmetric slab geometries, $\theta$ remains quantized to the bulk value, and provides an odd-integer contribution to Eq.~(\ref{eq:thetaconductance}) when $\theta\text{ mod }2\pi = \pi$~\cite{wieder2018axion} -- this effect manifests in finite 3D systems with $\nu_{x,y,z}=0$ (\emph{i.e.} AXIs) via chiral hinge modes [Fig.~\ref{fig:scheme}(a)].  Therefore, given a fixed, $\mathcal{I}$-symmetric, $\mathbf{R}_i$-directed slab, and knowledge of $\nu_{i}$, changes in $\theta$ can be numerically extracted through Eq.~(\ref{eq:thetaconductance}).  For the Weyl-CDWs in this work, our choice of origin corresponds to a convention in which $\theta=0$ when $\mathcal{G}_{z}=\nu_{z}L_{z}/(Nc)$~\footnote{This convention relies on $N$ being odd to avoid slabs with fractional modulated cells.  In the SM, we consider a Weyl-CDW with $N=2$, which can only be cut into an $\mathcal{I}$-symmetric slab with a half-integer number of modulated cells.}.

As an example, we analyze Eq.~(\ref{eq:ham0}) with commensurate $Q=2\pi/3c$.  The tight-binding model in the rBZ exhibits the symmetry-based indicators $\{2|001\}$ ($\{0|001\}$) at $\phi=0$ ($\phi=\pi$) [Eq.~(\ref{eq:z4def})]  (see \cite{SM}).  In Fig.~\ref{fig:commensuratewilson}(a,b), we plot the $\mathbf{\hat{y}}$-normal surface Green's function and spectrum, which are identical for $\phi=0,\pi$.  Because surface Green's functions do not capture hinge states~\cite{wieder2018axion,TMDHOTI,HingeSM,hotis,schindler2018higher,DiracInsulator}~\footnote{Specifically, surface Green's functions capture 2D degeneracies (\emph{e.g.} surface Dirac cones and weak QAH chiral modes) that are protected by the symmetries of the 2D wallpaper groups~\cite{DiracInsulator}.  Conversely, 1D hinge states are protected by the lower-symmetry quasi-1D frieze groups~\cite{WiederLayers,MTQC,hotis,ConwaySymmetries}.  When a higher-order TI hinge with a particular Miller index is smoothed into a surface, the surface Green's function will hence only detect surface states that can be stabilized by the symmetries of the exposed surface~\cite{hotis,song2017,hsu2019facetBismuth,DiracInsulator}.  Conversely, hinge states \emph{can} be detected in \emph{hinge} Green's function calculations, as employed in~\cite{HingeSM,xu2019higher}.} (or origin-dependent changes in $\theta$), then the spectral flow in Fig.~\ref{fig:commensuratewilson}(b) and the flat band in Fig.~\ref{fig:commensuratewilson}(a) indicate a bulk QAH contribution $\nu_{z}(\phi=0,\pi)=-1$.

\begin{figure}[t]
\includegraphics[width=0.8\columnwidth]{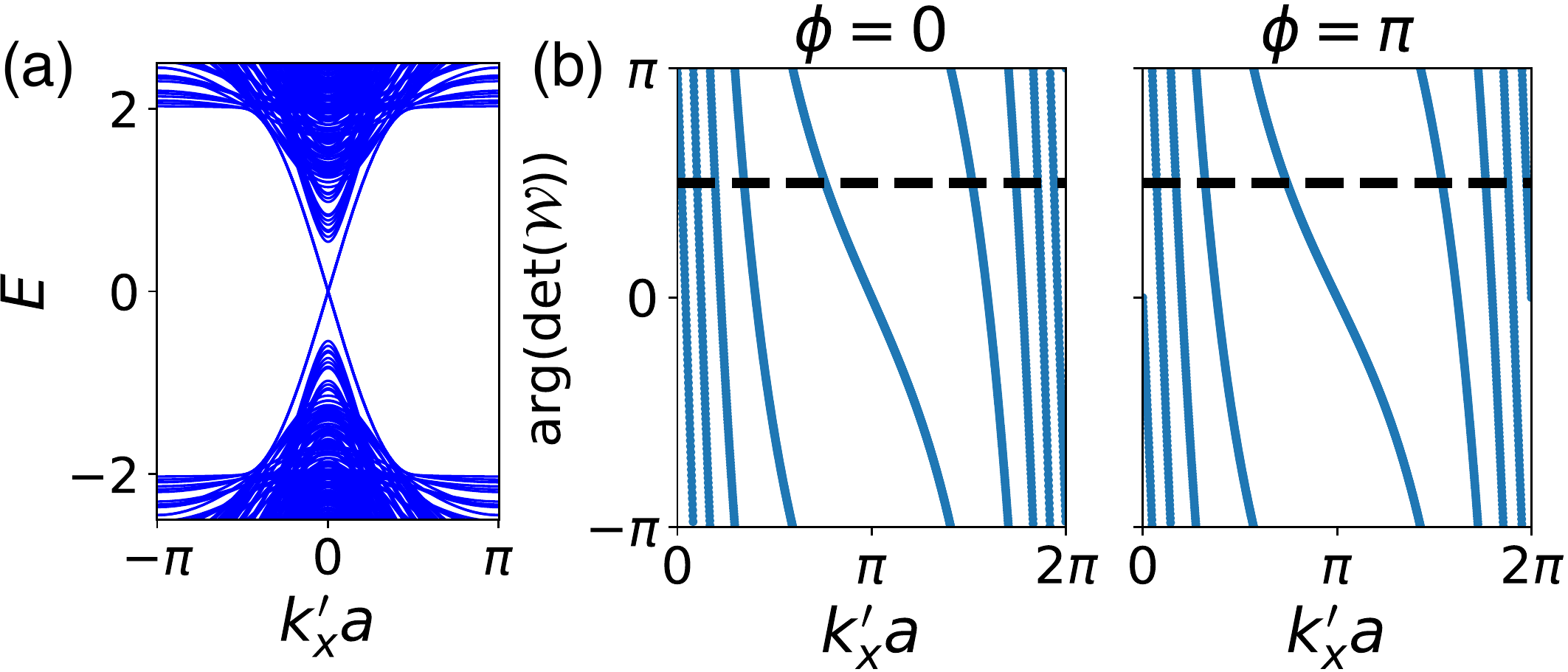}
\caption{(a) Band structure for the incommensurate Weyl-CDW in an $\mathcal{I}$-symmetric, $\hat{\bf x}$-directed rod geometry with 21 sites in the $\hat{\bf y}$ and $\hat{\bf z}$ directions and with $\phi=0$. The chiral states traversing the gap have degeneracy proportional to the rod thickness, indicating that they are QAH background surface states.  When $\phi=\pi$, the rod spectrum is qualitatively the same as (a), but exhibits one fewer pairs of chiral modes.  (b) The $\hat{\bf y}$-directed Wilson loop of an $\mathcal{I}$-symmetric, $\hat{\bf z}$-directed slab~\cite{wieder2018axion} of the model in (a) with 21 layers exhibits $\mathcal{G}_{z}=-9$ ($\mathcal{G}_{z}=-8$) spectral flow for $\phi=0$ ($\phi=\pi$), implying that $\delta\theta_{\phi}=\pi$.}
\label{fig:goldenrod}
\end{figure}

We next cut the model in Fig.~\ref{fig:commensuratewilson}(a,b) into an $\mathcal{I}$-symmetric, $\hat{\bf z}$-directed slab geometry with $L_{z}/3c=5$ unit cells, and calculate the $\hat{\bf y}$-directed non-abelian slab Berry phase (Wilson loop) $\mathcal{W}$~\cite{fidkowski2011bulk,Yu11,Alexandradinata16,wieder2018axion} [Fig.~\ref{fig:commensuratewilson}(c)], whose winding indicates that $\mathcal{G}_{z}(0)=-5=\nu_zL_{z}/3c,\ \mathcal{G}_{z}(\pi)=-4=\nu_zL_{z}/3c+1$.  Along with $\nu_{z}(\phi=0,\pi)=-1$, $|\delta \mathcal{G}_{z}| = 1$ indicates through Eq.~(\ref{eq:thetaconductance}) that the insulating Weyl-CDW at $\phi=0$ ($\phi=\pi$) is a QAH (oQAH) insulator, implying that $\delta\theta_{\phi}=\pi$.

Having demonstrated that $\mathcal{I}$-symmetric, commensurate Weyl-CDWs are either QAH or oQAH insulators, we next explore the case of incommensurate modulation.  Although an incommensurate CDW is not translationally-invariant, neither QAH nor oQAH phases require translation symmetry~\cite{qi2008topological,essin2009magnetoelectric,wieder2018axion,Hughes11,Turner2012,AFMTIMoore}.  Consequently, Eq.~(\ref{eq:thetaconductance}) still applies, without modification, to $\hat{\bf z}$-directed slabs of Eq.~(\ref{eq:ham0}) with incommensurate values of $Q$.

To confirm this result, we first cut $H_0+H_{CDW}$ [Eqs.~(\ref{eq:ham0}) and~(\ref{eq:cdwmass})] with $Q=\varphi\pi/2c$ (where $\varphi$ is the golden ratio) into an $\mathcal{I}$-symmetric rod geometry.  We observe an extensive number of QAH surface states along the rod for $\phi=0,\pi$, where there is exactly one fewer surface chiral mode at $\phi=\pi$ [Fig.~\ref{fig:goldenrod}(a)].  Next, to measure $\theta$, we cut the incommensurate Weyl-CDW into an $\mathcal{I}$-symmetric, $\hat{\bf z}$-directed slab and calculate the $\hat{\bf y}$-directed Berry phase, as we previously did in Fig.~\ref{fig:commensuratewilson}(c).  In the slab geometry, $|\delta \mathcal{G}_{z}|=1$ between $\phi =0,\pi$ [Fig.~\ref{fig:goldenrod}(b)].  Furthermore, in incommensurate CDWs, tuning $\phi$ changes the bulk wavefunctions, but not the bulk energy spectrum~\cite{kraus2012topological,de1983electrons}, such that a bulk-insulating, incommensurate Weyl-CDW with $\phi=0$ remains insulating at arbitrary $\phi$.  Additionally, $\nu_{z}$ cannot change without a bulk gap closure, whereas $\theta$ is free to wind between $0$ and $\pi$ at $\mathcal{I}$-breaking CDW angles away from $\phi=0,\pi$~\cite{qi2008topological,essin2009magnetoelectric,wieder2018axion}.  As shown in the SM, $|\delta \mathcal{G}_{z}|=1$ in Fig.~\ref{fig:goldenrod}(b), along with $G_{H,z}$ calculated for successive rational appoximants of an irrational $Q$, imply that, as in the commensurate case (Fig.~\ref{fig:commensuratewilson}), the incommensurate Weyl-CDW carries the relative axion angle $\delta\theta_{\phi}=\pi$.

Our results have several implications for experimental investigations of axionic response in Weyl-CDWs.  First, we have demonstrated that a large QAH effect is unavoidable and guaranteed in both commensurate and incommensurate minimal Weyl-CDWs, independent of $\phi$.  Second, the interplay between lattice and phase-angle defects, which both bind 1D chiral modes, is a fruitful area for future study, though one must cautiously separate contributions from $\theta$ and those from a background QAH effect~\cite{queiroz2019partial,frankprep}. Next, we emphasize that the axionic response in Weyl-CDWs is measurable through the dynamical dependence of the quasi-2D QAH effect on $\phi$, rather than through the static magnetoelectric polarizability at fixed $\phi$~\cite{armitage2019matter,lin2020prep}.  Furthermore, soliton-like defects in $\phi$ -- which carry the same half-quantized Hall conductivity as gapped AXI surfaces for $\delta \theta_{\phi} = \pi$~\cite{BurkovCDWWeyl} -- can in principle be manipulated by exciting the CDW sliding mode.  Finally, in $\mathcal{I}$-symmetric, magnetic Weyl-CDWs, our results highlight the experimental and theoretical difficulty of distinguishing QAH, oQAH, and AXI phases.  However, our results do imply that, by carefully computing ${\bf k}$-space Chern numbers and then zone-folding, it is possible to predict the topology of Weyl-CDWs in real materials without performing intensive quasiperiodic calculations.

Our methodology can be extended to spin-density waves~\cite{maciejko2014weyl,VladCDW,BeenakkerKramersWeyl,VTe2CDW}, and CDWs in $\mathcal{T}$-symmetric semimetals, including Dirac~\cite{Young12,ZJDirac,ZJDirac2}, Weyl, and nodal-line semimetals~\cite{Kim2015,fang2015topological,WiederLayers,LeslieLine1}, which also exhibit signatures of higher-order topology~\cite{TMDHOTI,HingeSM,TaylorToy,HingeDiracExp}.  Most interestingly, because rotation- and $\mathcal{T}$-symmetric HOTIs~\cite{wieder2018axion,fang2019new} can be formed from weak stacks of 2D TIs~\cite{Po2017,khalaf}, then rotation-symmetric CDWs in $\mathcal{T}$-symmetric WSMs, such as the CDW in (TaSe$_4$)$_2$I~\cite{shi2019charge}, may also exhibit nontrivial response effects.  Specifically, a CDW can fold four WPs in a rotation- and $\mathcal{T}$-symmetric WSM into an eightfold double Dirac point (DDP) in which line defects bind \emph{helical} modes equivalent to HOTI hinge states~\cite{Wieder2016b,Bradlyn2016}; alternatively a DDP critical point can also be realized by coupling two fourfold Dirac points with a CDW.  Recent experiments have demonstrated hinge-state-like step-edge helical modes and robust edge supercurrents in rotation- and $\mathcal{T}$-symmetric WSMs~\cite{DavidMoTe2Exp,MazWTe2Exp,WTe2HingeStep,PhuanOngMoTe2Hinge,MoTe2WeylInterface}, as well as a stable DDP and topological step-edge modes in the CDW phase of TaTe$_4$~\cite{BinghaiCDWDDP}.  In the SM, we present an explicit model demonstrating that a $\mathcal{T}$-symmetric Dirac-CDW hosts an eightfold DDP critical point that separates weak TI (WTI) and ``obstructed'' WTI (oWTI) phases that differ by a helical HOTI.  Unlike in the QAH and oQAH Weyl-CDW phases, the difference between the WTI and oWTI Dirac-CDWs cannot be connected to a known response theory, because a $\theta$-like topological field theory for helical HOTIs has not yet been elucidated~\cite{MTQC,wieder2018axion,TMDHOTI}, suggesting an intriguing direction for future study.

\begin{acknowledgments}
We acknowledge helpful conversations with Peter Abbamonte, B. Andrei Bernevig, Matthew Gilbert, Chao-Xing Liu, and Jiabin Yu, as well as crucial input from Zhida Song on an early version of this work.  B. J. W. acknowledges support from B. Andrei Bernevig through Department of Energy Grant No. DE-SC0016239, Simons Investigator Grant No. 404513, BSF Israel US Foundation Grant No. 2018226, and ONR Grant No. N00014-20-1-2303.  B. B. acknowledges the support of the Alfred P. Sloan Foundation, and the National Science Foundation Grant No. DMR-1945058. Concurrently with the revision of this work, Ref.~\cite{yu2020dynamical} also demonstrated that minimal $\mathcal{T}$-symmetric Weyl-CDWs are topologically equivalent to $\phi$-dependent weak TIs -- the results of Ref.~\cite{yu2020dynamical} are complementary to and in complete agreement with the results of this work. After the submission of this work, a stable DDP and topological step-edge modes were experimentally measured in the CDW phase of the $\mathcal{T}$-symmetric Dirac semimetal TaTe$_4$~\cite{BinghaiCDWDDP}, providing further support for the analysis performed in this work.  Additionally, after the submission of this work an analysis of minimal Weyl-CDWs beyond mean-field theory was performed in~\cite{BurkovCDWWeyl}; the analysis in~\cite{BurkovCDWWeyl} explicitly confirms our characterization of the mean-field QAH and oQAH phases of Weyl-CDWs.  Lastly, after the submission of this work, Ref.~\cite{LatestIvoHingePump} also demonstrated a relationship between AXI pumping cycles and hybrid-Wannier-sheet flow that is equivalent to the Weyl-CDWs studied in this work when the CDW angle $\phi$ is treated as a dynamical parameter.
\end{acknowledgments}
\bibliography{refs}
\end{document}


\title{Supplementary Material for ``Axionic Band Topology in Inversion-Symmetric Weyl-Charge-Density Waves''}
\author{Benjamin J. Wieder}
\affiliation{Department of Physics, Massachusetts Institute of Technology, Cambridge, MA 02139, USA}
\affiliation{Department of Physics, Northeastern University, Boston, MA 02115, USA}
\affiliation{Department of Physics, Princeton University, Princeton, New Jersey 08544, USA}

\author{Kuan-Sen Lin}
\author{Barry Bradlyn}
\affiliation{Department of Physics and Institute for Condensed Matter Theory, University of Illinois at Urbana-Champaign, Urbana, IL, 61801-3080, USA}

\date{\today}
\maketitle
\section{Parity (Inversion) Eigenvalue Analysis of the Minimal Magnetic Weyl Semimetal Tight-Binding Model without Modulation}
\label{sec:inveignnofold}

In this section, we compute the parity [inversion ($\mathcal{I}$)] eigenvalues of the occupied bands of the unperturbed tight-binding model with two Weyl points (WPs), which is described by the Bloch Hamiltonian:
\begin{equation}
    H_0(\mathbf{k})= -2(t_x\sigma^x\sin k_xa-t_y\sigma^y\sin k_yb)
     +2t_z\sigma^{z}(\cos k_zc -\cos\frac{Qc}{2})
    -m\sigma^{z}(2-\cos k_xa - \cos k_yb).
\label{eq:H0}
\end{equation}
Eq.~(\ref{eq:H0}) respects $\mathcal{I}$ symmetry, which is represented by:
\begin{equation}
\mathcal{I}H_{0}({\bf k})\mathcal{I}^{-1} = \sigma^{z}H_{0}(-{\bf k})\sigma^{z}.
\label{eq:Idef}
\end{equation}

\begin{table}[h]
\begin{tabular}{|c|c|c|}
\hline 
$(k_xa,k_yb,k_zc)$ & $n_+^a$ & $n_-^a$ \\
\hline 
\hline
$(0,0,0)$ & 0 & 1 \\
\hline
$(\pi,0,0)$ & 1 & 0 \\
\hline
$(0,\pi,0)$ & 1 & 0 \\
\hline
$(\pi,\pi,0)$ & 1 & 0 \\
\hline
$(0,0,\pi)$ & 1 & 0 \\
\hline
$(\pi,0,\pi)$ & 1 & 0 \\
\hline
$(0,\pi,\pi)$ & 1 & 0 \\
\hline
$(\pi,\pi,\pi)$ & 1 & 0 \\
\hline
\end{tabular}
\caption{Valence parity [inversion ($\mathcal{I}$)] eigenvalue multiplicities ($n_{\pm}^a$) for the unmodulated Hamiltonian $H_0({\bf k})$ of a minimal $\mathcal{I}$-symmetric Weyl semimetal [Eq.~(\ref{eq:H0})].}
\label{tab:inveigs}
\end{table}

We will now deduce the parity eigenvalues in the $ck_{z}=0,\pi$ planes.  First, at $k_zc=0$, Eq.~(\ref{eq:H0}) at the $\mathcal{I}$-invariant momenta (TRIM points) is given by:
\begin{align}
    H_0(0,0,0)&=2t_z(1-\cos\frac{Qc}{2})\sigma^z,\\ 
    H_0(\pi/a,\pi/b,0)&=-\left[4m-2t_z(1-\cos\frac{Qc}{2})\right]\sigma^z, \\
    H_0(\pi/a,0,0)&=H_0(0,\pi/b,0)=-\left[2m-2t_z(1-\cos\frac{Qc}{2})\right]\sigma^z.
\end{align}
We observe that $H_{0}(0,0,0)$, $H_{0}(\pi/a,0,0)$, $H_{0}(0,\pi/b,0)$, and $H(\pi/a,\pi/b,0)$ are proportional to the matrix $\mathcal{I}= \sigma^{z}$ [Eq.~(\ref{eq:Idef})], implying that, for $m\geq 2t_{z}$, the occupied parity eigenvalues at ${\bf k} = (\pi/a,0,0), (0,\pi/b,0)$, and $(\pi/a,\pi/b,0)$ are positive, but the occupied parity eigenvalue at ${\bf k}=(0,0,0)$ is negative.  Using the $\mathbb{Z}_{2}$ parity index for an $\mathcal{I}$-symmetric, time-reversal- ($\mathcal{T}$-) broken 2D insulator~\cite{QWZ,Hughes11,Fu2007,Turner2012}, we then compute the Chern number $C(k_zc)$ of the occupied bands in the $k_zc=0$ plane, which we find to be $C(k_{z}c=0)\text{ mod }2=-1$.  Conversely, in the $k_zc=\pi$ plane:
\begin{align}
    H_0(0,0,\pi/c)&=-2t_z(1+\cos\frac{Qc}{2})\sigma^z,\\ 
    H_0(\pi/a,\pi/b,\pi/c)&=-\left[4m+2t_z(1+\cos\frac{Qc}{2})\right]\sigma^z, \\
    H_0(\pi/a,0,\pi/c)&=H_0(0,\pi/b,\pi/c)= -\left[2m+2t_z(1+\cos\frac{Qc}{2})\right]\sigma^z,
\end{align}
implying that all of the parity eigenvalues of the occupied bands are positive, such that $C(k_{z}c=\pi)\text{ mod }2=0$.  Because $C(k_{z}c=\pi) - C(k_{z}c=0)\text{ mod }2=1$, then there must be a set of WPs with net chiral charge $|C| \mod 2 =1$ in each half of the Brillouin zone (BZ)~\cite{Wan11}, taking BZ halves to be indexed by positive and negative values of $k_{z}$.  We have verified that Eq.~(\ref{eq:H0}) only features two total WPs with $|C|=1$ chiral charges~\cite{mccormick2017minimal}.

We summarize the parity eigenvalues in Table~\ref{tab:inveigs}; we will find this information useful for future calculations (see Sec.~\ref{sec:inveigs}).

\section{Symmetry Indicators for $N$-Fold Commensurate Modulation}
\label{sec:inveigs}

We will now extend the previous analysis in Sec.~\ref{sec:inveignnofold} to the more general case of commensurate, $\mathcal{I}$-symmetric CDWs characterized by $Q=2\pi/Nc$ in Eq.~(\ref{eq:H0}).  This is accomplished by folding the BZ into the reduced BZ (rBZ) -- in which we label crystal momenta ${\bf k}'$ -- and then counting the parity eigenvalues that have been folded onto each TRIM point in the rBZ.  In the $Nk_z'c=0$ plane in the rBZ, the occupied Bloch states have been folded from all of the BZ planes at $k_zc=2\pi m/N,\ m\in\mathbb{Z}$ in the larger BZ of the unmodulated structure.  Crucially, we recognize that pairs of Bloch states folded from values of $k_{z}$ away from $k_{z}c=0,\pi$ will carry net-zero parity eigenvalues, because generic ${\bf k}$ points away from $k_{z}c=0,\pi$ are not $\mathcal{I}$-symmetric~\cite{Cracknell}.  Conversely, Bloch states folded from TRIM points in the original (unmodulated) cell will contribute the parity eigenvalues listed in Table~\ref{tab:inveigs}.

\begin{table}[h]
\begin{tabular}{|c|c|c|}
\hline
$(k'_xa,k'_yb,k'_zNc)$ & $n_+^a$ & $n_-^a$ \\
\hline
\hline
$(0,0,0)$ & $N/2$ & $N/2$ \\
\hline
$(\pi,0,0)$ & $(N/2)+1$& $(N/2)-1$ \\
\hline
$(0,\pi,0)$ & $(N/2)+1$ & $(N/2)-1$ \\
\hline
$(\pi,\pi,0)$ & $(N/2)+1$ & $(N/2)-1$ \\
\hline
$(0,0,\pi)$ & $[(N/2)-1]+\mathrm{Dirac}$ & $[(N/2)-1]+\mathrm{Dirac}$ \\
\hline
$(\pi,0,\pi)$ & $(N/2)$ & $(N/2)$ \\
\hline
$(0,\pi,\pi)$ & $(N/2)$ & $(N/2)$ \\
\hline
$(\pi,\pi,\pi)$ & $(N/2)$ & $(N/2)$ \\
\hline
\end{tabular}
\caption{Valence parity eigenvalue multiplicities ($n_{\pm}^a$) for the Hamiltonian of an $\mathcal{I}$-symmetric Weyl-CDW with $N$-fold modulation $Q=2\pi/Nc$ in Eq.~(\ref{eq:H0}), in the case in which $N$ is even.  The symbol $\mathrm{Dirac}$ represents the parity eigenvalue contribution from gapping the fourfold Dirac degeneracy that forms at the rBZ boundary (see the discussion in the main text).}
\label{tab:Neveneigs}
\end{table}

For $Q=2\pi/Nc$ modulation with an even value of $N$, the $k_zc=0,\pi$ planes and $N-2$ planes at generic values of $k_{z}$ from the original BZ are folded into the $k_{z}'Nc=0$ plane in the rBZ.  Conversely, the occupied bands at $k_{z}'Nc=\pi$ in the rBZ are folded from the $k_{z}$-indexed planes containing the two WPs in the original BZ, as well as $N-2$ planes at generic values of $k_{z}$.  It is important to note that, at the rBZ TRIM point ${\bf k}' = (0,0,\pi/Nc)$, the two WPs from the BZ have become folded into a fourfold Dirac degeneracy~\cite{Young12} in the rBZ.  As discussed in the main text, there is only one $\mathcal{I}$-symmetric mass for the Dirac degeneracy if it directly gaps, such that the Dirac point either contributes two negative or two positive parity eigenvalues to the set of valence parity eigenvalues.   The resulting distribution of valence parity eigenvalues is listed in Table~\ref{tab:Neveneigs}.

\begin{table}[h]
\begin{tabular}{|c|c|c|}
\hline
$(k'_xa,k'_yb,Nk'_zc)$ & $n_+^a$ & $n_-^a$ \\
\hline
\hline
$(0,0,0)$ & $(N-1)/2$ & $(N+1)/2$ \\
\hline
$(\pi,0,0)$ & $(N+1)/2$ & $(N-1)/2$ \\
\hline
$(0,\pi,0)$ & $(N+1)/2$ & $(N-1)/2$ \\
\hline
$(\pi,\pi,0)$ & $(N+1)/2$ & $(N-1)/2$ \\
\hline
$(0,0,\pi)$ & $[(N-1)/2]+\mathrm{Dirac}$ & $[(N-3)/2]+\mathrm{Dirac}$ \\
\hline
$(\pi,0,\pi)$ & $(N+1)/2$ & $(N-1)/2$ \\
\hline
$(0,\pi,\pi)$ & $(N+1)/2$ & $(N-1)/2$ \\
\hline
$(\pi,\pi,\pi)$ & $(N+1)/2$ & $(N-1)/2$ \\
\hline
\end{tabular}
\caption{Valence parity eigenvalue multiplicities ($n_{\pm}^a$) for the Hamiltonian of an $\mathcal{I}$-symmetric Weyl-CDW with $N$-fold modulation $Q=2\pi/Nc$ in Eq.~(\ref{eq:H0}), in the case in which $N$ is odd.  The symbol $\mathrm{Dirac}$ represents the parity eigenvalue contribution from gapping the fourfold Dirac degeneracy that forms at the rBZ boundary (see the discussion in the main text).}
\label{tab:Noddeigs}
\end{table}

For $Q=2\pi/Nc$ modulation with an odd value of $N$, the situation is similar to the even case.  In the $k_z'Nc=0$ plane of the rBZ, there are folded Bloch states originating from the $k_zc=0$ plane in the BZ of the unmodulated structure, as well as Bloch states from $N-1$ additional planes at generic values of $k_{z}$.  In the $Nck'_z=\pi$ plane of the rBZ, there are states originating from the $k_zc=\pi$ plane of the original BZ, Bloch states from the two $k_{z}$-indexed planes containing WPs in the original BZ, and Bloch states originating from $N-3$ additional planes at generic values of $k_{z}$.  As in the even case, at the rBZ TRIM point ${\bf k}' = (0,0,\pi/Nc)$, the two WPs from the BZ have become folded into a fourfold Dirac degeneracy~\cite{Young12} in the rBZ.  The resulting distribution of valence parity eigenvalues is listed in Table~\ref{tab:Noddeigs}.

For all $N\in \mathbb{Z}^{+}$, when the Dirac degeneracy at $k_{z}'Nc=\pi$ is directly gapped, the resulting occupied parity eigenvalues are determined by the CDW phase $\phi$.    As determined through the ${\bf k}\cdot {\bf p}$ analysis in the main text, when $\phi=0$ ($\phi=\pi$), the Dirac degeneracy is split to contribute two additional negative (positive) parity eigenvalues.  Because the Weyl-CDW with commensurate $N$ has 3D translation and $\mathcal{I}$ symmetries, then it respects the symmetries of magnetic space group MSG 2.4 ($P\bar{1}$), numbered in the convention of Belov, Nerenova, and Smirnova (BNS)~\cite{BelovNotation,Cracknell} (the Hamiltonian also respects additional symmetries that do not affect the analysis performed in this work).  Previous works~\cite{xu2020high,NaturePaper,MTQC,Po2017,watanabe2018structure,khalaf2018higher,wieder2018axion,JiabinZhidaAXIDirac,MurakamiAXI1,MurakamiAXI2} have determined the symmetry-based indicators in MSG 2.4 ($P\bar{1}$) to be given by:
\begin{equation}
    \tilde{z}_4\equiv\frac{1}{2}\sum_{\mathbf{k}_a\in\mathrm{TRIMS}}(n^a_+-n^a_-) \mod 4,
\end{equation}
and:
\begin{equation}
    \tilde{z}_{2i}\equiv \frac{1}{2}\sum_{\substack{\mathbf{k}_a\in\mathrm{TRIMS}\\ \mathbf{k}_a\cdot\mathbf{R}_i=\pi}}(n^a_+-n^a_-) \mod 2,
\end{equation}
where in particular, odd values of weak $\tilde{z}_{2i}$, along with the presence of a bulk gap at all ${\bf k}$ points (\emph{i.e.}, the absence of WPs), indicate that the Chern number $C_{i}(k_{i})\text{ mod }2=1$ at all values of $k_{i}$.  Using the parity eigenvalues listed in Tables~\ref{tab:Neveneigs} and~\ref{tab:Noddeigs}, we deduce that:
\begin{align}
    \tilde{z}_4(\phi=n\pi)= 1+(-1)^n,\qquad \tilde{z}_{2x}=\tilde{z}_{2y}=0,\qquad \tilde{z}_{2z}=1.
    \label{eq:finalStrongWeak}
\end{align}
In particular, the nontrivial, $\phi$-independent weak Chern index $\tilde{z}_{2z}=1$ directly indicates the presence of an unavoidable QAH background with an odd Chern number, as discussed throughout the main text.

\section{Extension to Incommensurate Modulation}

In this section, we will show that the previous zone-folding arguments in Sec.~\ref{sec:inveigs} can be extended beyond integer values of $N$ in $Q=2\pi/Nc$.  To begin, in the case in which $Q=2\pi M/Nc$, where $M$ and $N$ are coprime positive integers, the rBZ continues to have a well-defined primitive reciprocal lattice vector $\mathbf{G}_z=(2\pi/Nc)\hat{\bf z}$.  However, when $Q=2\pi M/Nc$, the two WPs are separated by a momentum $N\delta k'_zc = 2\pi M$. This  implies that $M$ planes indexed by $k_{z}$ with $|k_{z}|< \delta k'_{z}/2$ in the original BZ -- whose occupied bands each carry a Chern number $\nu_{z}=-1$ -- will be folded onto the same value of $k_{z}'$ in the rBZ.  Thus, in the gapped Weyl-CDW phase with $Q=2\pi M/Nc$, there is a weak Chern number $\nu_{z}=-M$.  When $M$ is odd, the two WPs in the original BZ continue to fold onto the ${\bf k}'=(0,0,\pi/Nc)$ point in the rBZ, and the previous parity eigenvalue analysis in Sec.~\ref{sec:inveigs} applies.  However, when $M$ is even, the two WPs fold onto the ${\bf k}'={\bf 0}$ point in the rBZ.  Nevertheless, the ${\bf k}\cdot {\bf p}$ Dirac Hamiltonian that we analyzed in the main text is agnostic to the TRIM point around which it is formulated, and therefore also characterizes the fourfold Dirac point that forms at ${\bf k}'={\bf 0}$ when $|\Delta|=0$ and $M$ is even.  Therefore, the dependence of the strong ($\tilde{z}_{4}$) and weak $(\tilde{z}_{2i}$) indices on the CDW phase $\phi$ [Eq.~(\ref{eq:finalStrongWeak})] continues to imply that, for a $\hat{\bf z}$-directed slab with thickness $L_{z}$ of a Weyl-CDW with modulation $Q= 2\pi M / Nc$, the anomalous Hall conductance $G_{H,z}$ is given by:
\begin{equation}
    G_{H,z}=\frac{e^2}{h}\left(-\frac{ML_{z}}{Nc}+\frac{\theta_\phi}{\pi}\right),
\end{equation}
where $\delta\theta_\phi=\theta_{\phi=\pi}-\theta_{\phi=0} \mod 2\pi=\pi$ relative to a fixed choice of origin (see the main text for a more general expression for slab Hall conductance).

Crucially, because any irrational modulation can be expressed as the limit of a sequence of rational approximants~\cite{rudin1964principles}, this implies that our results also extend to Weyl-CDWs with incommensurate modulation.  Taking the limit in which $N/M$ becomes irrational while keeping the slab Chern number $ML_z/Nc$ fixed, we conclude that, for both commensurate and incommensurate Weyl-CDWs formed from minimal $\mathcal{I}$-symmetric Weyl semimetals, the anomalous Hall conductance of a $\hat{\bf z}$-directed slab with thickness $L_{z}$ is given by:
\begin{equation}
G_{H,z}=\frac{e^2}{2\pi h} \left(-QL_{z}+2\theta_\phi\right).
\label{eq:irrationalAndrationalGh}
\end{equation}
Because $\delta\theta_{\phi}=\pi$ for every rational $Q$, then we conclude that $\delta\theta_{\phi}=\pi$ remains true in the limit of irrational $Q$. To leading order in the thermodynamic limit, Eq.~(\ref{eq:irrationalAndrationalGh}) is consistent with the statement that the CDW preserves the anomalous Hall conductance obtained by integrating the ${\bf k}$-space Chern numbers between the WPs of the parent (unmodulated) Weyl semimetal.

\section{Explicit Model of a Commensurate Weyl-CDW with $N=2$ Modulation}
\label{sec:n2}

In the case of $Q= 2\pi/Nc = \pi/c$ ($N=2$) modulation (\emph{i.e.}, a Peierls distortion), the unmodulated Hamiltonian $H_{0}$ in Eq.~(\ref{eq:H0}) can simply be re-expressed in a larger unit that is doubled in the $\hat{\bf{z}}$- ($c$-axis-) direction.  The position-space embedding of the $N=2$ modulated Hamiltonian then arises from four total orbitals, which are distributed into pairs located at $(x,y,z)=(0,0,0)$ and $(0,0,c)$ within the doubled cell.  Introducing a vector of Pauli matrices $\vec{\mu}$ to index the orbitals at $z=0,c$, we obtain the unmodulated Hamiltonian:
\begin{align}
H_0(\mathbf{k'})&=2(-t_x\sigma^x\mu^0\sin k'_xa +t_y\sigma^y\mu^0\sin k'_yb +t_z\cos {k'_zc}\sigma^z\mu^x) \nonumber \\
    &-m(2-\cos k'_xa - \cos k'_yb)\sigma^z\mu^0,
\label{eq:N2mod}
\end{align} 
where ${\bf k}'$ indexes momentum in the reduced BZ (rBZ).  In the rBZ, the reciprocal lattice vectors are given by ${\bf G}_{x}' = (2\pi/a)\hat{\bf x}$, ${\bf G}_{y}' = (2\pi/b)\hat{\bf y}$, and ${\bf G}_{z}' = (\pi/c)\hat{\bf z}$, such that $H_0(2ck_{z}' + 2\pi)$ is now related to $H_0(2k_z'c)$ by the reciprocal lattice vector ${\bf G}_{z}'$.  Eq.~(\ref{eq:N2mod}) is $\mathcal{I}$ symmetric; using induction from the real-space data~\cite{EBRTheoryPaper} (\emph{i.e.}, the positions of the four orbitals), we deduce that the matrix representative of $\mathcal{I}$ at each of the eight rBZ TRIM points $\mathbf{k'}_a$ is given by:
\begin{equation}
    \mathcal{I}(\mathbf{k}'_a)=\sigma^z\otimes\left(\begin{array}{cc}
    1 & 0 \\
    0 & e^{-i2ck'_{az}}
    \end{array}\right).
  \label{eq:N2iDef}
\end{equation}

We next consider the modulation induced by the CDW distortion, which is given by $H_{CDW}$ in the main text.  For the case in which $N=2$, $QR_z\in\pi\mathbb{Z}$, such that:
\begin{equation}
    \cos (QR_z+\phi) = (-1)^{\frac{QR_z}{\pi}}\cos\phi.
\label{eq:N2phi}
\end{equation}
Eq.~(\ref{eq:N2phi}) implies that the on-site modulation $H_{CDW}$ assumes a form:
\begin{equation}
    H_{CDW}=|\Delta|\cos\phi\sigma^z\mu^z.
    \label{eq:N2cdw}
\end{equation}
Surprisingly, for $N=2$ (Peierls) modulation, $H_{CDW}$ is $\mathcal{I}$-symmetric at \emph{all} values of $\phi$, distinctly unlike the more generic CDWs discussed throughout this work.  Because $\mathcal{I} = \sigma^{z}\mu^{z}$ at the four TRIM points in the $2ck'_z=\pi$ plane [Eq.~(\ref{eq:N2iDef})], then the ${\bf k}\cdot {\bf p}$ expansion of $H_{0} + H_{CDW}$ around ${\bf k}' = (0,0,\pi/2c)$ acquires a Dirac mass $V_{\phi}=\mathcal{I}|\Delta|\cos\phi$.  Consequently, for $|\Delta|> 0$, the Dirac degeneracy at ${\bf k}'=(0,0,\pi /2c)$ splits, leading at half filling to two occupied bands at ${\bf k}'=(0,0,\pi /2c)$ with negative [positive] parity ($\mathcal{I}$) eigenvalues when $\sgn[\cos\phi]$ is positive [negative].

For values of $|\Delta|$ small enough to not invert bands elsewhere in the rBZ, the occupied parity eigenvalues at the four TRIM points in the $2ck'_z=0$ plane are given by $\{+-,++,++,++\}$ for all values of $\phi$.  Conversely, the valence parity eigenvalues in the $2ck'_z=\pi$ plane are $\phi$-dependent, and are given by  $\{--,+-,+-,+-\}$ when $\sgn[\cos \phi]$ is positive and by $\{++,+-,+-,+-\}$ when $\sgn[\cos \phi]$ is negative (unlike in the case of general modulation, for $N=2$ modulation $Q=\pi/c$, the bulk is gapless at $\phi = \pm \pi/ 2$).  We observe that the occupied parity eigenvalues computed for $Q=\pi/c$ ($N=2$) modulation are consistent with the valence parity eigenvalue formulas for even-$N$ modulation listed in Table~\ref{tab:Neveneigs}.  Using the symmetry-based indicators $\{\tilde{z}_4|\tilde{z}_{2x}\tilde{z}_{2y}\tilde{z}_{2z}\}$ for 3D crystals with $\mathcal{I}$ and translation symmetry (see Sec.~\ref{sec:inveigs} and the main text)~\cite{xu2020high,NaturePaper,MTQC,Po2017,watanabe2018structure,khalaf2018higher,wieder2018axion,JiabinZhidaAXIDirac,MurakamiAXI1,MurakamiAXI2}, we determine that the occupied bands of the $N=2$ minimal Weyl-CDW are characterized by the indices $\{2|001\}$ [$\{0|001\}$] when $\sgn[\cos \phi]$ is positive [negative].  From the slab Berry phase analysis performed in the main text, we conclude [using the origin choice specified by the embedding in Eq.~(\ref{eq:N2mod})] that the Weyl-CDW with $N=2$ modulation is a weak Chern (QAH) insulator with $\nu_{z}=-1$, $\theta=\nu_{x}=\nu_{y}=0$ when $\sgn[\cos \phi]$ is positive, and consequently is an ``obstructed'' QAH (oQAH) insulator with $\theta=\pi$, $\nu_{z}=-1$, $\nu_{x}=\nu_{y}=0$ when $\sgn[\cos \phi]$ is negative.

In the four-band model $H_0+H_{CDW}$, we find that $\delta\theta_\phi=\theta_{\phi=\pi} - \theta_{\phi=0}\text{ mod }2\pi$ can in fact be directly computed.  At nonzero values of $k_{x}'$ and $k_{y}'$, the Hamiltonian $H_{0}+H_{CDW}$ is generically gapped for all values of $\phi$, allowing us to specialize momentarily to the line $k_{x}'=k_{y}'=0$ along which the bulk gap is smallest (or vanishing for $\phi = \pm \pi/2$):
\begin{equation}
    H(0,0,k_z',\phi)=\sigma^z(2t_z\mu^x\cos k'_zc+\mu^z|\Delta|\cos\phi).
\end{equation}
First, we fix $\phi$ to a value away from $\phi =\pm \pi/2$; this opens a bulk gap at all ${\bf k}'$ points in the rBZ.  Next, we tune $t_{z}$ to zero, which \emph{does not} close a bulk gap.  With $t_{z}\rightarrow 0$, $[H_{0}+H_{CDW},\mu^{z}]=0$, allowing us to divide the two occupied bands into $\mu^{z}=\pm 1$ sectors.  When $\phi$ is fixed to a value $\phi\neq \pm\pi/2$ and then $t_{z}$ is set to zero without closing the bulk gap, $H_{0}+H_{CDW}$ assumes the $k_{z}'$-independent form of a 2D Chern insulator~\cite{bernevigbook}, but \emph{only within one of the $\mu^{z}=\pm 1$ sectors} [Eqs.~(\ref{eq:N2mod}) and~(\ref{eq:N2cdw})]; in the other ($\mu^{z}=\mp 1$) sector, the occupied band carries a trivial Chern number.

In the case in which $\cos\phi>0$, the band with $\mu^{z}=1$ [$\mu^{z}=-1$] in $H_{0}+H_{CDW}$ carries the Chern number $C_{z}(k_{z}')=-1$ [$C_{z}(k_{z}')=0$] for all values of $k_{z}'$ in the rBZ.  In the definition of $\mathcal{I}$ symmetry in Eq.~(\ref{eq:N2iDef}), the $\mu^{z}=1$ ($\mu^{z}=-1$) sector is $k_{z}'$-independent ($k_{z}'$-dependent).  First, we note that because the $\mu^{z}=1$ subspace of $H_{0}+H_{CDW}$ is $k_{z}'$-independent, and has a $k_{z}'$-independent embedding [from the definition of $\mathcal{I}$ in Eq.~(\ref{eq:N2iDef})], then we can inverse-Fourier-transform the $k_{z}'$ component of the $\mu^{z}=1$ subspace of $H_{0}+H_{CDW}$ to realize a $C_{z}=-1$ Chern insulator at $z=0$.  Next, we recognize that, while the $\mu^{z}=-1$ subspace in the $\mathcal{I}$ embedding is $k_{z}'$-dependent [Eq.~(\ref{eq:N2iDef})], the occupied band in the $\mu^{z}=-1$ subspace is topologically trivial.  We have confirmed that the trivial occupied band in the $\mu^{z}=-1$ subspace of $H_{0}+H_{CDW}$ with $\cos\phi>0$ can be inverse-Fourier-transformed in 3D into a maximally- (exponentially-) localized Wannier orbital~\cite{Marzari2012} at $(x,y,z) = (0,0,c)$.  Therefore, consistent with our earlier determination in this section that the Weyl-CDW with $N=2$ modulation and phase $\cos\phi >0$ carries the topological indices $\nu_{z}=-1$, $\theta=\nu_{x}=\nu_{y}=0$, we find that  $H_{0}+H_{CDW}$ with $\cos\phi >0$ can be deformed into the limit of a weak stack~\cite{song2017,xu2020high,NaturePaper,MTQC,MurakamiAXI1,MurakamiAXI2} of one $C_{z}=-1$ Chern insulator in the $z=0$ plane plus one Wannier orbital at $(x,y,z)=(0,0,c)$ per doubled cell.

In the case in which  $\cos\phi<0$, we find that the resulting orbitals and Chern insulators are related to the previous case of $\cos\phi > 0$ by a (now-fractional) lattice translation ${\bf t} = c\hat{\bf z}$, similar to the relationship between the two phases of the Su-Schrieffer-Heeger (SSH) model of polyacetylene~\cite{ssh1979} (\emph{c.f.}~\cite{queiroz2019partial}).   To see this, we first note that when $\cos\phi < 0$, the band with $\mu^{z}=1$ [$\mu^{z}=-1$] in $H_{0}+H_{CDW}$ carries the Chern number $C_{z}(k_{z}')=0$ [$C_{z}(k_{z}')=-1$] for all values of $k_{z}'$ in the rBZ.  We then again note that, in the definition of $\mathcal{I}$ symmetry in Eq.~(\ref{eq:N2iDef}), the $\mu^{z}=1$ ($\mu^{z}=-1$) sector is $k_{z}'$-independent ($k_{z}'$-dependent).  This implies that a trivial band in the $\mu^{z}=1$ sector can be inverse-Fourier-transformed into a Wannier orbital at $(x,y,z)=(0,0,0)$, but also implies a more complicated dependence on $k_{z}'$ for the $C_{z}(k_{z}')=-1$ band in the $\mu^{z}=-1$ sector.  However, if we shift our origin by an amount ${\bf t} = c\hat{\bf z}$, we restore the previous relationship between the embedding in $\mathcal{I}$ and the $\mu^{z}=\pm 1$ sectors of the ${\bf k}'$-space Hamiltonian $H_{0}+H_{CDW}$; specifically, we again realize a situation in which the band with nontrivial Chern number lies in the same $\mu^{z}$ sector as the sector of $\mathcal{I}$ with no $k_{z}'$ dependence.  This allows us to once again inverse-Fourier-transform the $k_{z}'$ components of the two occupied bands of $H_{0}+H_{CDW}$ with $\cos\phi<0$ to realize a weak stack of Wannier orbitals and Chern insulators.  However, in the case of $\cos\phi<0$, the weak stack consists of one Wannier orbital at $(x,y,z)=(0,0,0)$ plus one $C_{z}=-1$ Chern insulator in the $z=c$ plane, which we recognize as an oQAH insulator with the topological indices $\theta=\pi$, $\nu_{z}=-1$, $\nu_{x}=\nu_{y}=0$.  The case of $N=2$ modulation thus emphasizes the origin-dependence of $\theta$ when $\nu_{x,y,z}\neq 0$.  However, in the presence of $\mathcal{I}$ symmetry, the difference $\delta\theta_{\phi}=\pi$ between $\cos\phi>0$ and $\cos\phi<0$ is origin- and gauge-independent and topological, analogous to the difference in the positions of the Wannier centers between the two phases of the SSH model~\cite{ssh1979,vanderbiltaxion,zilberberg2018photonic}.

We can also analytically determine that $\delta\theta_{\phi} = \pi$ for an $\mathcal{I}$-symmetric Weyl-CDW with $N=2$ modulation by appealing to the field-theoretic discussion in the main text and expressing $\theta$ in terms of the non-abelian Berry connection $\mathcal{A}$.  While the Chern-Simons form $CS_3[\mathcal{A}]$ appearing in the integrand of Eq.~(2) of the main text can be a bit cumbersome to evaluate due to gauge-fixing in the presence of nonzero weak Chern numbers, we can simplify the analysis by computing the difference $\delta\theta_{\phi}$~\cite{qi2008topological,essin2009magnetoelectric}:
\begin{align}
    \delta\theta_{\phi}& = \theta_{\phi=\pi}-\theta_{\phi=0}\text{ mod }2\pi=\frac{1}{4\pi}\int CS_3[\mathcal{A}(\phi=\pi)]- CS_3[\mathcal{A}(\phi=0)] \\
    &=\frac{1}{16\pi}\int d^3k' dk_4 \epsilon^{\mu\nu\lambda\rho}\mathrm{tr}(\Omega_{\mu\nu}\Omega_{\lambda\rho}), 
    \label{eq:pontryagin}
\end{align}
where $\Omega$ is the non-abelian Berry \emph{curvature} for the occupied states of the Hamiltonian:
\begin{equation}
    H(\mathbf{k}',k_4)=H_0(\mathbf{k}') + k_4\sigma^z\mu^z,
\end{equation}
and where the integral in Eq.~(\ref{eq:pontryagin}) is taken within the range $k_4=-|\Delta|$ to $k_4=|\Delta|$.  $H(\mathbf{k}',k_4)$ is $\mathcal{I}$-symmetric, where we define $\mathcal{I}$-symmetry as leaving $k_4$ invariant (since $H_{CDW}$ is $\mathcal{I}$-even).  The presence of $\mathcal{I}$ symmetry causes the contribution to the integral in Eq.~(\ref{eq:pontryagin}) to be zero from every $\mathcal{I}$-symmetric region that does not enclose a gapless point of $H(\mathbf{k}',k_4)$ at half filling.  Thus, we can deform the region of integration to an infinitesimal ball surrounding the degeneracy at $(\mathbf{k}',k_4)=(0,0,\pi/2c,0)$.  We next expand the Hamiltonian $H(\mathbf{k}',k_4)$ about the degenerate point in $(\mathbf{k}',k_4)=(0,0,\pi/2c,0)+(\mathbf{q'},q_4)$, and obtain the ${\bf k}\cdot {\bf p}$ Hamiltonian of a 4D fourfold Dirac fermion:  
\begin{equation}
H(\mathbf{q}',q_{4},\Delta)\approx -t_xaq'_x\sigma^x\mu^0 + t_ybq_y'\sigma^y\mu^0 + t_zc q_z'\sigma^z\mu^x + q_4 \sigma^z\mu^z,
\label{eq:4Dkp}
\end{equation}
For the Dirac Hamiltonian in Eq.~(\ref{eq:4Dkp}), the integrand $\epsilon^{\mu\nu\lambda\rho}\mathrm{tr}(\Omega_{\mu\nu}\Omega_{\lambda\rho})=16\pi^2\delta(\mathbf{q},q_4)$ in Eq.~(\ref{eq:pontryagin}) takes the form of a point source~\cite{ryu2010topological,qi2008topological}, allowing the integral to be computed directly:
\begin{equation}
  \delta\theta_{\phi} =   \theta_{\phi=\pi}-\theta_{\phi=0}\text{ mod }2\pi = \pi.
\end{equation}
Having previously established that $\nu_{z}=-1$, $\nu_{x}=\nu_{y}=0$ for $H_{0}+H_{CDW}$ when $\cos\phi<1$, we conclude that the $N=2$ Weyl-CDW with phase $\cos\phi<1$ is an oQAH insulator with the symmetry-based indicators $\{0|001\}$ and the topological indices $\theta=\pi$, $\nu_{z}=-1$, $\nu_{x}=\nu_{y}=0$, in agreement with more general formulation in Sec.~\ref{sec:inveigs}, and consistent with our analysis of the other Weyl-CDW models in this work [though we again note that the relative assignment of $\theta$ depends on the choice of origin, which we have here chosen to be $(x,y,z)=(0,0,0)$].

\section{Weyl-CDWs with Nontrivial Chern Numbers at $k_z=\pi$}

In this section, we consider a slightly distinct Weyl semimetal (WSM) from the model studied in the main text.  Here, we again consider a WSM with $\mathcal{I}$ and 3D translation symmetries, but one in which the $k_{z}=\pi/c$ plane now carries a nontrivial Chern number $C(k_{z}c=\pi)=1$, instead of the $k_{z}=0$ plane.  To realize this configuration of Chern numbers [$C(k_zc=0)=0,\ C(k_zc=\pi)=1$], we reverse the signs of $t_x,\ t_y$, and $m$ in $H_{0}$ in the main text.   We again introduce the on-site CDW mass $H_{CDW}$ with a commensurate modulation in the $\hat{\bf z}$-direction by $Q=2\pi/Nc$, and then again fold bands into the rBZ.  For the WSM in this section, each band \emph{outside} of the rBZ now contributes $+1$ to the weak Chern number, because each plane outside of the rBZ has the ${\bf k}$-space Chern number $C(|k_zc|>\pi/N)=1$.  In this case, independent of the CDW phase $\phi$, we find that the gapped Weyl-CDW phase still unavoidably carries the weak Chern numbers $\nu_x=\nu_y=0$, $\nu_z=N-1$.

We next determine the parity eigenvalues and symmetry-based indicators~\cite{xu2020high,NaturePaper,MTQC,Po2017,watanabe2018structure,khalaf2018higher,wieder2018axion,JiabinZhidaAXIDirac,MurakamiAXI1,MurakamiAXI2} of the WSM in the presence of a CDW with commensurate modulation $Q=2\pi/Nc$.  For even values of $N$, the resulting band topology is the same as in the previous analysis performed in Sec.~\ref{sec:inveigs}.  However, when $N$ is odd, the weak symmetry index $\tilde{z}_{2z}= N -1\text{ mod }2=0$, implying that $\nu_{z}$ is \emph{even} (though we have shown in the previous paragraph that $\nu_{z}$ cannot be zero).  Next, we determine the strong index $\tilde{z}_{4}$ by repeating the analysis used to construct Table~\ref{tab:Noddeigs}.  Through this analysis, we find that there are $2N+2$ positive and $2N-2$ negative valence parity eigenvalues in the $k_z'Nc=0$ plane, and $2N-2$ negative and $2N$ positive parity eigenvalues and one fourfold Dirac fermion in the $k_z'Nc=\pi$ plane.  Computing the strong index $\tilde{z}_4$, we find that:
\begin{equation}
    \tilde{z}_4(\phi=n\pi)=1-(-1)^n.
\end{equation}
Thus, when $\phi=0$ the Weyl-CDW, in this section is a $\nu_{z}=N-1$ weak Chern insulator with the \emph{completely trivial} parity indices $\{0|000\}$, and when $\phi=\pi$, the Weyl-CDW is an oQAH insulator with the  parity indices $\{2|000\}$.  Crucially, from the analysis performed in this section and throughout this work, we determine that the oQAH phase of the Weyl-CDW carries the topological indices $\nu_{z}=N-1$, $\nu_{x}=\nu_{y}=0$, and that the Weyl-CDWs at $\phi =0,\pi$ differ by a topological axion angle $\delta\theta_{\phi}=\pi$.

\begin{figure}[h]
\includegraphics[width=0.5\textwidth]{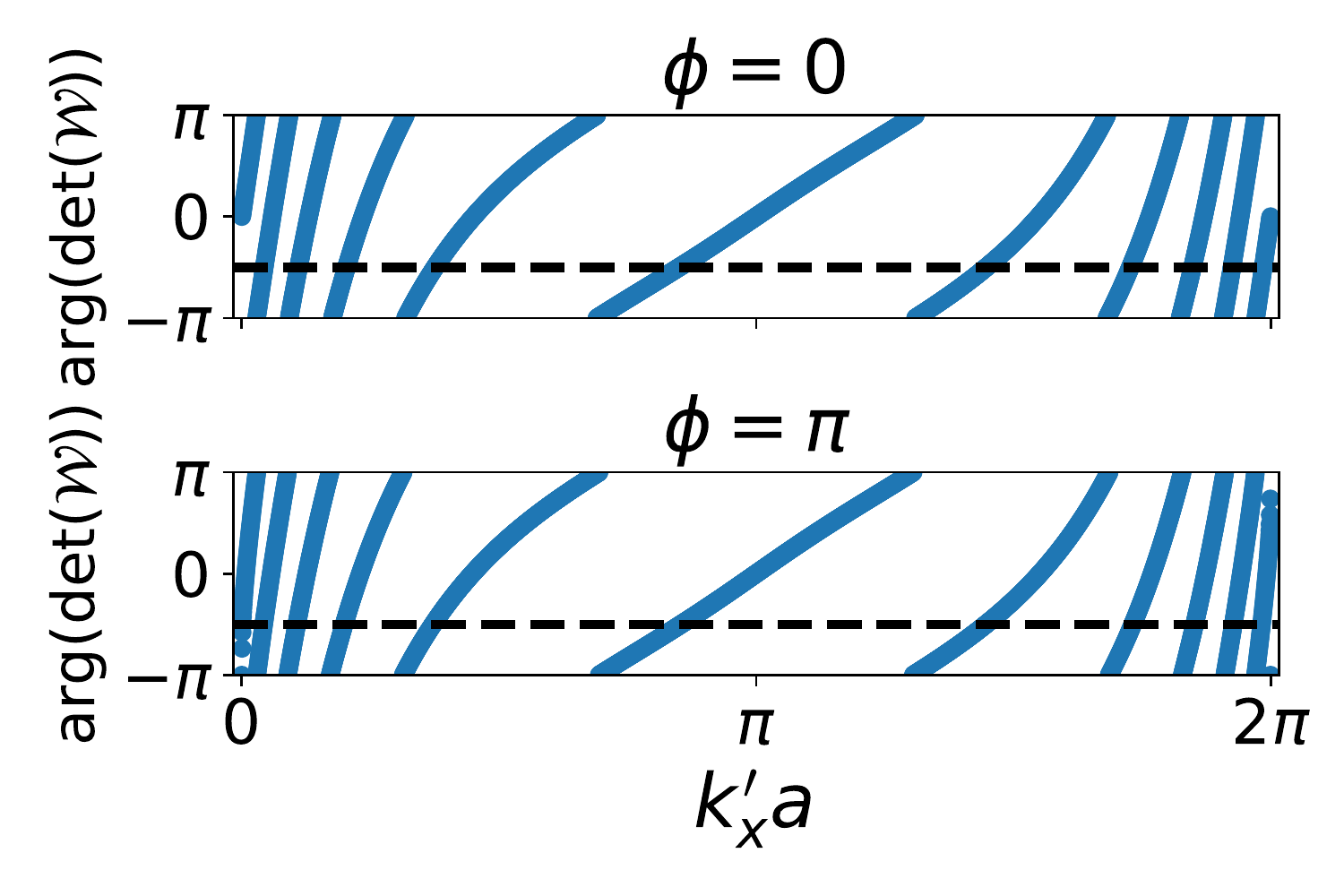}
\caption{$\hat{\bf y}$-directed Berry phase of an $\mathcal{I}$-symmetric, $\hat{\bf z}$-directed slab with $L_{z}/3c=5$ layers of the $N=3$ Weyl-CDW discussed in this section, for which we have determined that the weak Chern number is $\nu_{z}=2$, independent of the CDW phase $\phi$.  When $\phi=0$, the Berry phase crosses the dashed line with a positive slope $L_{z}\nu_z/3c=10$ times, indicating that $\theta_{\phi=0}=0$ in the convention (\emph{i.e.} choice of origin) used throughout this work.  Conversely, when $\phi=\pi$, the Berry phase crosses the dashed line with a positive slope $(L_{z}\nu_{z}/3c) + 1 = 11$ times, indicating that $\theta_{\phi=\pi}=\pi$.  Thus, the Weyl-CDWs at $\phi=0,\pi$ differ by a topological axion angle $\delta\theta_{\phi}=\pi$.}
\label{fig:slab}
\end{figure}

To confirm the presence of a nontrivial axion angle at $\theta=\pi$ (in the origin-dependent convention employed throughout this work), we perform numerical calculations for the case of commensurate modulation with $N=3$.  In Fig.~\ref{fig:slab}, we have calculated the $\hat{\bf y}$-directed Berry phase of an $\mathcal{I}$-symmetric, $\hat{\bf z}$-directed slab with $15$ atomic layers, corresponding to $L_{z}/3c=5$ unit cells of the modulated Weyl-CDW phase.  When $\phi=0$, the Berry phase implies a slab Chern number $\mathcal{G}_{z}(\phi=0)=10$, which is consistent with a weak Chern number $\nu_{z} = N-1= 2$ through $\mathcal{G}_{z} = L_{z}\nu_{z}/3c$.  Conversely, when $\phi=\pi$, the Berry phase in Fig.~\ref{fig:slab} implies a slab Chern number of $\mathcal{G}_{z}(\phi = \pi) = 11$.  This is consistent with the observation that $|\delta\tilde{z}_4|=2$ indicates a difference in the parity of the Chern number of an $\mathcal{I}$-symmetric slab~\cite{MurakamiAXI1,MurakamiAXI2,xu2020high,Turner2012,Hughes11}. As discussed in the main text, for a fixed, $\mathcal{I}$-symmetric slab geometry with more than one layer, $|\delta \mathcal{G}_{z}|=1$ implies that $\delta \theta_{\phi}= \pi$.  From the zone-folding analysis performed earlier in this section and the discussion in the main text, Fig.~\ref{fig:slab} implies that, given our choice of origin, the Weyl-CDW is an oQAH insulator with $\theta=\pi$, $\nu_{z}=2$, $\nu_{x}=\nu_{y}=0$.  Crucially, the difference $\delta\theta_{\phi}= \theta_{\phi=\pi}- \theta_{\phi=0}\text{ mod }2\pi=\pi$ is \emph{origin-independent} and topological.

\section{Time-Reversal-Invariant Semimetal-CDWs}

In this section, we will show how our earlier analysis of magnetic Weyl-CDWs can be generalized to $\mathcal{T}$-invariant CDW phases of topological semimetals.  In particular, we will show how a CDW distortion in an $\mathcal{I}$- and $\mathcal{T}$-symmetric Dirac semimetal can open a topological gap, yielding two topologically distinct weak topological insulator (WTI) phases.  We begin with a $\mathcal{T}$-doubled extension of Eq.~(\ref{eq:H0}):
\begin{equation}
    H({\bf k})=\tau^z\left[m\left(\cos k_xa + \cos k_yb -2\right) + 2t_z\left(\cos k_zc - \cos \frac{Qc}{2}\right)\right]-2t_x \tau^x\sigma^z\sin k_xa +2t_y \tau^y\sin k_yb,
\label{eq:TDiracCDW}
\end{equation}
where $\vec{\tau}$ are a set of Pauli matrices that act in orbital space, and $\vec{\sigma}$ are a set of Pauli matrices that act in spin space [we have relabeled the orbital Pauli matrices from $\sigma$ in the $\mathcal{T}$-breaking Weyl-CDWs elsewhere in this work to $\tau$ in Eq.~(\ref{eq:TDiracCDW}) to draw connection with the Bernevig-Hughes-Zhang and Fu-Kane-Mele models of 2D and 3D topological insulators (TIs)~\cite{qi2008topological,bernevig2006quantum,Kane04,kane2005z,fukanemele,Fu2007}].  Eq.~(\ref{eq:TDiracCDW}) is symmetric under $\mathcal{I}$ and $\mathcal{T}$, which are respectively represented by:
\begin{equation}
\mathcal{I}H({\bf k})\mathcal{I}^{-1} = \tau^{z}H(-{\bf k})\tau^{z},\ \mathcal{T}H({\bf k})\mathcal{T}^{-1} = \sigma^{y}H^{*}(-{\bf k})\sigma^{y}.
\label{eq:symmetriesWithTSymmetry}
\end{equation}
$H({\bf k})$ in Eq.~(\ref{eq:TDiracCDW}) is gapped except for two fourfold-degenerate Dirac points at $\mathbf{k}=(0,0,\pm Qc/2)$, and has two negative energy (occupied, valence) bands, and two positive energy (unoccupied, conduction) bands.   We note that, because a fourfold Dirac degeneracy may be considered the superposition of two $|C|=1$ WPs with opposite chiral charges~\cite{MurakamiWeylDiracPhase}, then Eq.~(\ref{eq:TDiracCDW}) may also be considered an unstable intermediate point during the onset of a bidirectional CDW~\cite{BidirectionalKivelson}.  Specifically, we may consider the Dirac points in Eq.~(\ref{eq:TDiracCDW}) to be formed from a CDW that couples and folds the four WPs in a minimal $\mathcal{T}$-symmetric, $\mathcal{I}$-broken WSM into two fourfold Dirac points.  In this (more unrealistic) picture, the CDW that folds WPs into Dirac points also shifts atoms to induce a structural transition to an intermediate, centro- ($\mathcal{I}$-) symmetric, metastable structural phase.  When we subsequently couple the two Dirac points with a $\hat{\bf z}$-directed CDW (see below), the CDW that couples the Dirac points may thus be considered the second contribution to a bidirectional CDW in a minimal $\mathcal{T}$-symmetric WSM.  Though it is unrealistic to consider the case in which a bidirectional CDW restores centrosymmetry, because WTI phases can also be stabilized in noncentrosymmetric crystals~\cite{song2017}, then it is possible to generalize the construction below to more realistic models of minimal $\mathcal{T}$-symmetric Weyl-CDWs.  We leave this possibility for future works.

As previously in Sec.~\ref{sec:inveignnofold}, we can compute and analyze the valence parity eigenvalues in the $k_zc=0, \pi$ planes.  Evaluating $H({\bf k})$ at each of the eight TRIM points:
\begin{align}
    H(0,0,0)&=2t_z(1-\cos\frac{Qc}{2})\tau^z, \\
    H(\pi/a,\pi/b,0)&=[2t_z(1-\cos\frac{Qc}{2})-4m]\tau^z, \\
    H(\pi/a,0,0)&=H(0,\pi/b,0)=[2t_z(1-\cos\frac{Qc}{2})-2m]\tau^z, \\
    H(0,0,\pi/c)&=-2t_z(1+\cos\frac{Qc}{2})\tau^z, \\
    H(\pi/a,\pi/b,\pi/c)&=[-2t_z(1+\cos\frac{Qc}{2})-4m]\tau^z, \\
    H(\pi/a,0,\pi/c)&=H(0,\pi/b,\pi/c)=[-2t_z(1+\cos\frac{Qc}{2})-2m]\tau^z,
\label{eq:parityEvalsWithT}
\end{align}
we observe that $H({\bf k})$ is proportional to the matrix representative of $\mathcal{I}$ [Eq.~(\ref{eq:symmetriesWithTSymmetry})] at each TRIM point.  For $m\geq 2t_z>0$, the two occupied states at ${\bf k}=(0,0,0)$ have negative parity eigenvalues, while the two occupied states at each of the other seven TRIM points have positive parity eigenvalues (see Table~\ref{tab:Tinveigs}).  Using the Fu-Kane parity index for the 2D $\mathbb{Z}_2$ invariant $z_{2D}(k_{z}=0,\pi/c)$~\cite{Fu2007}, the parity eigenvalues in Table~\ref{tab:Tinveigs} imply that the $k_zc=0$ plane of $H({\bf k})$ is topologically equivalent to a 2D TI $z_{2D}(0)=1$, while the $k_zc=\pi$ plane of $H({\bf k})$ exhibits a trivial $\mathbb{Z}_{2}$ invariant $z_{2D}(\pi/c)=0$.  We note that, unlike in the case of the magnetic WSM phase analyzed in the other sections of this supplement and in the main text of this work, the difference in the $\mathbb{Z}_{2}$ indices $z_{2D}(\pi/c) - z_{2D}(0)\text{ mod }2=1$ does not necessarily imply the existence of Dirac points, because the $\mathbb{Z}_{2}$ invariant, unlike the 2D Chern number, is not well-defined in BZ planes without $\mathcal{T}$ symmetry [\emph{e.g.} $k_{z}\neq 0,\pi/c$].  This can be summarized by the statement that the momentum-space Chern number implies the existence of WPs through a descent relation, whereas the $\mathbb{Z}_{2}$ invariant does not universally imply the existence of fourfold Dirac fermions through an analogous descent relation (see~\cite{HingeSM}).  We also note that, because we have not explicitly enforced rotation symmetries, the Dirac points along $k_{x}=k_{y}=0$ in Eq.~(\ref{eq:TDiracCDW}) are generically unstable to $\mathcal{I}$- and $\mathcal{T}$-symmetric perturbations~\cite{ZJDirac,Yang2014}.  For completeness, we note that Eq.~(\ref{eq:TDiracCDW}) can be tuned to a $C_{4z}$-symmetric limit in which the Dirac points are symmetry-stabilized ($t_x=t_y, a=b$), however, our analysis below does not require the presence of $C_{4z}$ or other rotation symmetries.

\begin{table}[h]
\begin{tabular}{|c|c|c|}
\hline 
$(k_xa,k_yb,k_zc)$ & $n_+^a$ & $n_-^a$ \\
\hline 
\hline
$(0,0,0)$ & 0 & 2 \\
\hline
$(\pi,0,0)$ & 2 & 0 \\
\hline
$(0,\pi,0)$ & 2 & 0 \\
\hline
$(\pi,\pi,0)$ & 2 & 0 \\
\hline
$(0,0,\pi)$ & 2 & 0 \\
\hline
$(\pi,0,\pi)$ & 2 & 0 \\
\hline
$(0,\pi,\pi)$ & 2 & 0 \\
\hline
$(\pi,\pi,\pi)$ & 2 & 0 \\
\hline
\end{tabular}
\caption{Valence parity [inversion ($\mathcal{I}$)] eigenvalue multiplicities ($n_{\pm}^a$) for the unmodulated Hamiltonian $H({\bf k})$ of a minimal $\mathcal{I}$- and $\mathcal{T}$-symmetric Dirac semimetal [Eq.~(\ref{eq:TDiracCDW})].}
\label{tab:Tinveigs}
\end{table}

We will now analyze Eq.~(\ref{eq:TDiracCDW}) in the presence of a CDW.  To guarantee that the CDW that couples the bulk Dirac points induces a gap, we posit a $\mathcal{T}$-symmetric symmetry-lowering CDW order that explicitly breaks the $C_{4z}$ symmetry that is present in Eq.~(\ref{eq:TDiracCDW}) in the limit that $t_x=t_y, a=b$; in the mean-field, the CDW potential takes the form:
\begin{align}
    H_\Delta=\Delta\sum_\mathbf{k}&\left[c^\dag_{\mathbf{k+Q}}\tau^z e^{i\phi} c_\mathbf{k} + c^\dag_{\mathbf{k-Q}}\tau^z e^{-i\phi} c_\mathbf{k} + c^\dag_\mathbf{k}\left[ v_1\tau^x\sigma^x(\sin k_zc) (\cos[k_xa] -1) + v_2\tau^x\sigma^x(\sin k_zc) (\cos[k_ya] -1)  \right]c_\mathbf{k}\right].
\label{eq:TCDW}
\end{align}
The first term in Eq.~(\ref{eq:TCDW}) is a translation-symmetry-breaking potential that opens a gap between the two Dirac points, while the remaining two terms represent mirror- and rotational-symmetry-breaking couplings that may be induced by the CDW order [for example, in the case of (TaSe$_4$)$_2$I, the CDW order breaks both translational and fourfold rotational symmetry~\cite{shi2019charge}].  We also note that, although the perturbation $H_\Delta$ contains spin-dependent terms, $H_\Delta$ respects $\mathcal{T}$ symmetry [Eq.~(\ref{eq:symmetriesWithTSymmetry})] and does not induce a non-zero spin expectation value at any position -- hence, $H_\Delta$ is indeed the mean-field potential of a (spinful) CDW, as opposed to a spin-density wave. Thus, the full Hamiltonian:
\begin{equation}
   \sum_\mathbf{k} c^\dag_\mathbf{k} H(\mathbf{k}) c_\mathbf{k} +H_\Delta,
\end{equation}
for the distorted system is $\mathcal{T}$-invariant for all $\phi$, and $\mathcal{I}$-symmetric for $\phi=0,\pi$.

In complete analogy with our previous analysis of $\mathcal{I}$-symmetric minimal Weyl-CDWs in Sec.~\ref{sec:inveigs}, when $Qc=2\pi/N$ in Eq.~(\ref{eq:TCDW}), we can use zone-folding arguments to analyze the Dirac-CDW phase.  In the rBZ, the two Dirac points are folded by a $Qc=2\pi/N$ CDW on top of one another, resulting in a (generically unstable) eightfold double Dirac point (DDP)~\cite{Wieder2016,Bradlyn2016} (though first-principles calculations and quantum oscillations experiments of the CDW phase of TaTe$_4$ have demonstrated that it is possible for a CDW to fold two fourfold Dirac fermions into a \emph{stable} eightfold DDP, given a highly-specific combination of CDW-order symmetries~\cite{BinghaiCDWDDP}).  In the Dirac-CDW model in this section, the eightfold DDP is explicitly gapped by $H_\Delta$.  We next compute the valence parity eigenvalues and $\mathcal{T}$-invariant symmetry-based indicators for the insulating Dirac-CDW phase in the $\mathcal{I}$-symmetric limits of Eq.~(\ref{eq:TCDW}) $\phi=0,\pi$.  The analysis can be simplified by observing that, because both Bloch states in every Kramers degeneracy at each TRIM point in the Dirac-CDW phase have the same parity eigenvalues, and that the parity eigenvalues in Table~\ref{tab:Tinveigs} are equivalent to two ($\mathcal{T}$-reversed copies) of the Weyl-CDW parity eigenvalues in Table~\ref{tab:inveigs}, then the counting of the parity of Kramers pairs of states in the rBZ is identical to the counting of the parity of states in Sec.~\ref{sec:inveigs}.  With $\mathcal{I}$- and $\mathcal{T}$-symmetry, the symmetry-based indicators can be obtained by subducing bands onto Shubnikov SG (SSG) 2.5 ($P\bar{1}1'$)~\cite{khalaf,Po2017,song2017,xu2020high,MTQC,TMDHOTI}.  In SSG 2.5 ($P\bar{1}1'$), the strong $\mathbb{Z}_{4}$ index is given by:
\begin{equation}
    z_4\equiv\frac{1}{4}\sum_{\mathbf{k}_a\in\mathrm{TRIMS}}(n^a_+-n^a_-)\text{ mod }4,
\end{equation}
and the weak $\mathbb{Z}_2$ indices are given by:
\begin{equation}
    z_{2i}\equiv \frac{1}{4}\sum_{\substack{\mathbf{k}_a\in\mathrm{TRIMS}\\ \mathbf{k}_a\cdot\mathbf{R}_i=\pi}}(n^a_+-n^a_-)\text{ mod }2.
\end{equation}
Thus, the previous analysis in Sec.~\ref{sec:inveigs} directly implies that, for the $\mathcal{T}$-symmetric Dirac-CDW phase analyzed in this section:
\begin{align}
    z_4(\phi=n\pi)= 1+(-1)^n,\qquad z_{2x}=z_{2y}=0,\qquad z_{2z}=1.
    \label{eq:TfinalStrongWeak}
\end{align}
Eq.~(\ref{eq:TfinalStrongWeak}) implies that when $\Delta\neq 0$, there is a nonzero weak TI index vector $(z_{2x},z_{2y},z_{2z})=(0,0,1)$, independent of $\phi$. In addition, the strong index $z_4$ switches between $z_4(\phi=0)=2$ at $\phi=0$ and $z_4(\phi=\pi)=0$ at $\phi=\pi$.  This shows that, in analogy with the minimal $\mathcal{I}$-symmetric Weyl-CDW considered in the main text, the eightfold DDP at $\Delta\rightarrow 0$ represents a $\delta z_{4,\phi}=z_{4}(\pi) - z_{4}(0)\text{ mod }4=2$ helical higher-order TI (HOTI) critical point [\emph{i.e.} two superposed, $\mathcal{T}$-reversed copies of a fourfold Dirac axion insulator (AXI) critical point], embedded in a background WTI vacuum [\emph{i.e.} two superposed, $\mathcal{T}$-reversed copies of a QAH background with odd $\nu_{z}$].

We will  now explore the physical consequences of the $\phi$-dependent topology of the Dirac-CDW.  First, we recognize that we can understand both the $\{0|001\}$ and $\{2|001\}$ phases in SSG 2.5 ($P\bar{1}1'$) from a layer-construction perspective. Following~\cite{song2017}, we note that a periodic array of $\hat{\bf z}$-normal 2D TIs stacked along the $z$-direction at $z=0$ in each unit cell carries the symmetry-based indicators $\{z_{4}|z_{2x}z_{2y}z_{2z}\}=\{2|001\}$; similarly, a $\hat{\bf z}$-directed stack of $\hat{\bf z}$-normal 2D TIs at $z=Nc/2$ in each unit cell carries the symmetry-based indicators $\{0|001\}$. Thus, in analogy with the main text, we may refer to the $\{2|001\}$ phase at $\phi=0$ as a weak TI (WTI), and the $\{0|001\}$ phase at $\phi=\pi$ as an ``obstructed'' WTI (oWTI).  Analogously to the QAH and oQAH phases analyzed in the main text, the WTI and oWTI phases of the Dirac-CDW also differ by a half-lattice translation in the modulated cell.  The difference between the QAH and oQAH phases is a $\{\tilde{z}_4|\tilde{z}_{2x}\tilde{z}_{2y}\tilde{z}_{2z}\}=\{2|000\}$ AXI with $\theta=\pi$ (see Fig.~1 of the main text).  Correspondingly, the difference between the WTI and the oWTI phases must be equivalent to two, superposed, $\mathcal{T}$-reversed AXIs, and is therefore a $\{z_{4}|z_{2x}z_{2y}z_{2z}\}=\{2|000\}$ helical HOTI~\cite{wieder2018axion,TMDHOTI}.  However, unlike in the case of the QAH and oQAH Weyl-CDWs, the difference between the WTI and oWTI Dirac-CDWs $\delta z_{4,\phi}=2$ cannot be connected to a known response theory, because a $\theta$-like topological field theory for helical HOTIs has not yet been elucidated.  We leave this exciting avenue of study for future works.

\begin{figure}[t]
\includegraphics[height=2in]{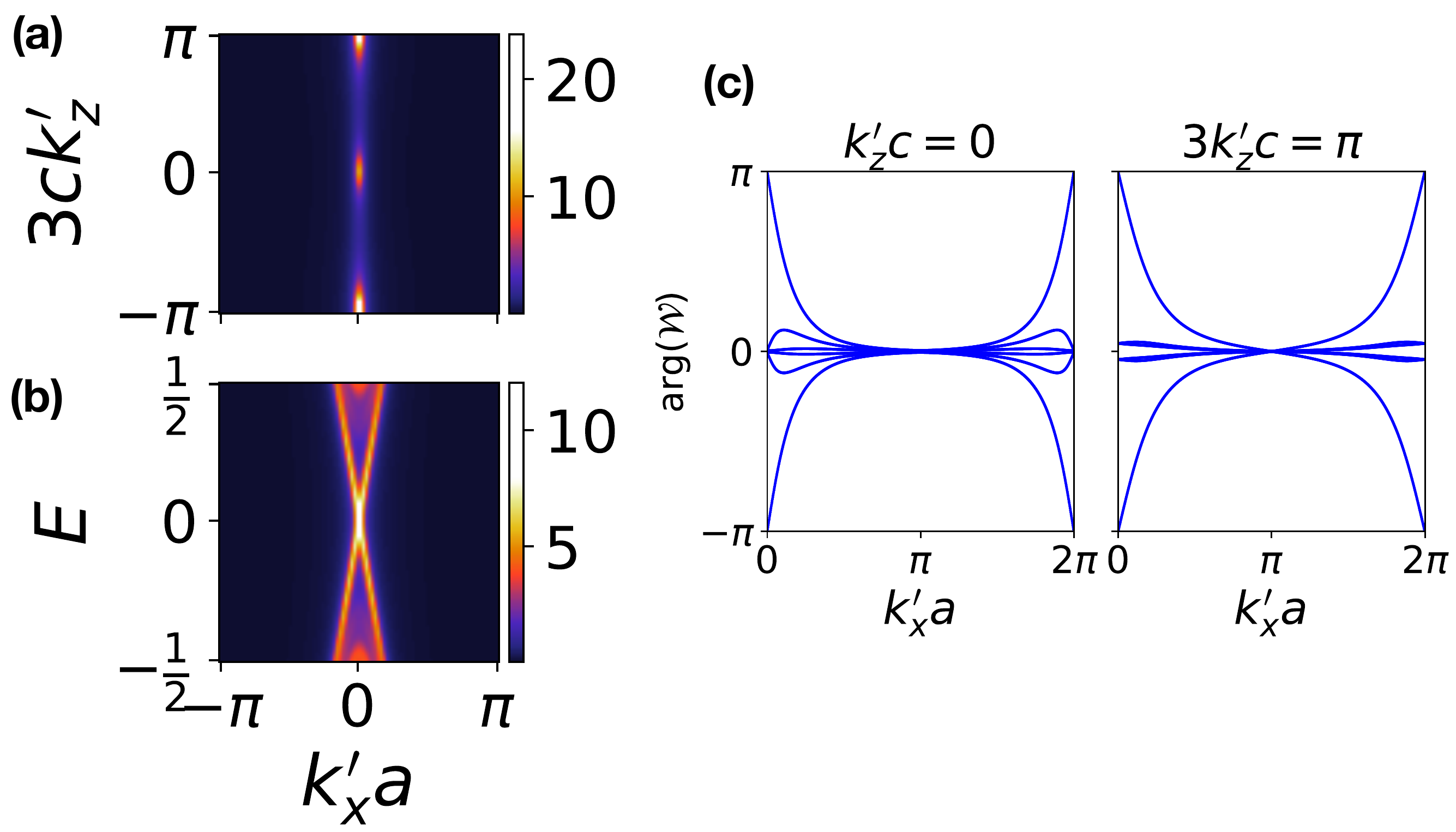}
\caption{Surface spectral function and Wilson loops for the $\mathcal{T}$-invariant Dirac-CDW with $Qc=2\pi/3$ at $\phi=0$. (a) The $\hat{\bf y}$-normal surface spectral function of a $\hat{\bf y}$-directed slab of the minimal $\mathcal{I}$- and $\mathcal{T}$-symmetric Dirac-CDW in Eqs.~(\ref{eq:TDiracCDW}) and~(\ref{eq:TCDW}) at energy $E=0$ as a function of $k_x'a$ and $3k_z'c$. Twofold degeneracies can be observed at $(k_x'a,3k_z'c)=(0,0),(0.\pi)$. (b) The surface spectral function of the slab in (a) at fixed $k_z'Nc=0$ as a function of $k_x'a$ and $E$.  The surface spectrum in (b) exhibits the characteristic linear dispersion of one of the two, twofold Dirac cones of the side surface of a WTI.  We have numerically confirmed that the twofold degeneracy at $(k_x'a,3k_z'c)=(0,\pi)$ in (a) is also a linearly-dispersing, twofold Dirac fermion.  (c) The bulk, $\hat{\bf y}$-directed Wilson loop as a function of $k_x'a$ at $3k_z'c=0,\pi$.  The helical winding of the Wilson loop eigenvalues confirm that the Dirac-CDW phase of Eqs.~(\ref{eq:TDiracCDW}) and~(\ref{eq:TCDW}) exhibits the side surface states and bulk spectral flow of a WTI (or obstructed WTI) Although these figures were generated using $\phi=0$, we have numerically confirmed that the surface spectral functions and bulk Wilson-loop winding are both $\phi$-independent.}
\label{fig:slabspec}
\end{figure}

To detect the difference between the WTI and oWTI phases, we can consider a finite, $\mathcal{I}$-symmetric slab of a Dirac-CDW.  Although the slab Hall conductance will identically vanish by $\mathcal{T}$-symmetry, we expect that, analogously to the QAH and oQAH phases of the Weyl-CDW, as $\phi$ evolves from $0$ to $\pi$, the slab will transition from a 2D TI into a 2D insulator with a trivial $\mathbb{Z}_{2}$ index due to the transition between WTI and oWTI phases.  To validate this interpretation, we consider the case of $Qc=2\pi/3$ modulation and analyze the distorted model in the rBZ.  In Fig.~\ref{fig:slabspec}, we first show the $\hat{\bf y}$-normal surface spectral function for a $\hat{\bf y}$-directed Dirac-CDW slab with $\phi=0$ that is translationally invariant in the $\mathbf{\hat{x}}$- and $\mathbf{\hat{z}}$-directions.  In Fig.~\ref{fig:slabspec}(a), we plot the surface spectral function at zero energy as a function of $k_x'a$ and $3k_z'c$; a nonzero density of surface states can be observed at $3k_z'c=0$ and $3k_z'c=\pi$.  In Fig~\ref{fig:slabspec}(b), we plot the surface spectral function of the same slab at fixed $3k_z'c=0$ as a function of $k_x'a$ and energy, confirming that the surface degeneracies in Fig~\ref{fig:slabspec}(a) are two, twofold-degenerate Dirac-cone side-surface states. The surface spectra in Fig~\ref{fig:slabspec}(a,b) are consistent with a WTI (or oWTI) with a trivial Fu-Kane strong 3D index $z_4\text{ mod }2 = 0$.   This is further corroborated by bulk Wilson loops, as shown in Fig.~\ref{fig:slabspec}(c). Although the plots in Fig.~\ref{fig:slabspec} were generated at $\phi=0$, we observe topologically indistinguishable surface and Wilson-loop spectra for \emph{all} values of $\phi$.  This is consistent with the recognition that the momentum-space $\mathbb{Z}_{2}$ indices $z_{2D}(k_{z}'Nc=0,\pi)$ are quantized by $\mathcal{T}$ symmetry, even in the absence of $\mathcal{I}$ symmetry [although the bulk topology is only symmetry-indicated in the $\mathcal{I}$-symmetric limits $\phi=0,\pi$].

\begin{figure}[h]
\includegraphics[height=2in]{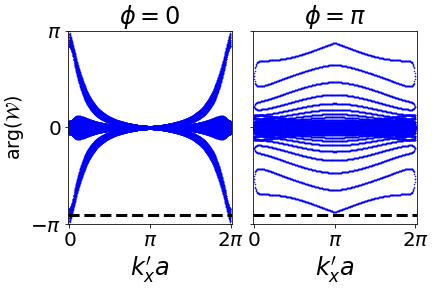}
\caption{ $\hat{\bf y}$-directed Wilson loop of an $\mathcal{I}$-symmetric, $\mathbf{\hat{z}}$-directed slab of a minimal $\mathcal{I}$- and $\mathcal{T}$-symmetric Dirac-CDW [Eqs.~(\ref{eq:TDiracCDW}) and~(\ref{eq:TCDW})] with $Qc=2\pi/3$ containing $L_z/3c=5$ (distorted) unit cells.  (a) The slab Wilson loop at $\phi=0$. The Wilson loop eigenvalues cross the black dashed line 15 times in each half of the rBZ, indicating that the Dirac-CDW slab at $\phi=0$ is a 2D TI, implying that the bulk Dirac-CDW is a WTI. (b) The Wilson loop of the slab in (a) with $\phi=\pi$. In (b), the Wilson loop eigenvalues do not cross the black dashed line, indicating that the Dirac-CDW slab at $\phi=\pi$ is has a trivial $\mathbb{Z}_{2}$ index, implying in combination with the surface spectral function and bulk Wilson loop calculations in Fig.~\ref{fig:slabspec} that the Dirac-CDW at $\phi=\pi$ is an obstructed WTI that is equivalent to two, $\mathcal{T}$-reversed copies of the oQAH phase in Fig.~1(c) of the main text.}
\label{fig:TRslabwilson}
\end{figure}

Lastly, in Fig.~\ref{fig:TRslabwilson}, we compute the $\hat{\bf y}$-directed Wilson loop of a $\mathbf{\hat{z}}$-directed Dirac-CDW slab with $L_z/3c = 5$ layers at $\phi=0$ and $\phi=\pi$ . When $\phi=0$, the Wilson loop eigenvalues helically wind -- specifically, there are $15$ Wilson bands crossing the horizontal dashed line in each half of the BZ in Fig.~\ref{fig:TRslabwilson} at $\phi=0$. This indicates that the slab as a whole can be viewed as a 2D TI when $\phi=0$~\cite{Yu11,Soluyanov2011}.  In contrast, when $\phi=\pi$, there is no winding in the slab Wilson spectrum, indicating that the slab exhibits a trivial $\mathbb{Z}_{2}$ invariant.  In conclusion, we have shown that, in a minimal $\mathcal{I}$- and $\mathcal{T}$-symmetric Dirac-CDW, the difference in the strong $\mathbb{Z}_{4}$ invariant $\delta z_{4,\phi} = 2$ may be detected in the slab Wilson loop, even though there does not yet exist a topological response theory associated to $z_{4}$ [as opposed to the strong index $\tilde{z}_{4}$ of magnetic AXIs and Weyl-CDWs, for which $\delta \tilde{z}_{4}=2$ carries the axionic response of a Chern-Simons 3-form, see Eq.~(\ref{eq:4Dkp}) and the surrounding text].

\bibliography{refs}